\documentclass[twocolumn]{aastex631}

\newcommand\aastex{AAS\TeX}
\newcommand\latex{La\TeX}
\newcommand\fobs{$f_{\rm obs}$}
\newcommand\fint{$f_{\rm int}$}
\newcommand\xqr{XQR-30}
\newcommand\xshoot{X-shooter}
\newcommand\kms{km s$^{-1}$}
\newcommand\bi{$BI$}
\newcommand\vmax{$v_{\rm max}$}
\newcommand\vmin{$v_{\rm min}$}
\newcommand\vlim{$v_{\rm lim}$}
\newcommand\wmax{$w_{\rm max}$}

\newcommand\lbol{$L_{\rm Bol}$}
\newcommand\ledd{$\lambda_{\rm Edd}$}

\usepackage{multirow}
\usepackage{hyperref}
\usepackage[utf8]{inputenc}

\shorttitle{BAL quasar fraction and kinematics across cosmic time}
\shortauthors{Bischetti et al.}

\begin{document}

\title{The fraction and kinematics of broad absorption line quasars across cosmic time}

\correspondingauthor{Manuela Bischetti}
\email{manuela.bischetti@inaf.it}

\author[0000-0002-4314-021X]{Manuela Bischetti}
\affiliation{Dipartimento di Fisica, Universit\'{a} di Trieste, Sezione di Astronomia, Via G.B. Tiepolo 11, I-34131 Trieste, Italy}
\affiliation{INAF - Osservatorio Astronomico di Trieste, Via G. B. Tiepolo 11, I--34143 Trieste, Italy }

\author[0000-0002-4031-4157]{Fabrizio Fiore}
\affiliation{INAF - Osservatorio Astronomico di Trieste, Via G. B. Tiepolo 11, I--34143 Trieste, Italy }
\affiliation{IFPU - Institut for fundamental physics of the Universe, Via Beirut 2, 34014 Trieste, Italy }

\author[0000-0002-4227-6035]{Chiara Feruglio}
\affiliation{INAF - Osservatorio Astronomico di Trieste, Via G. B. Tiepolo 11, I--34143 Trieste, Italy }
\affiliation{IFPU - Institut for fundamental physics of the Universe, Via Beirut 2, 34014 Trieste, Italy }

\author[0000-0003-3693-3091]{Valentina D'Odorico}
\affiliation{INAF - Osservatorio Astronomico di Trieste, Via G. B. Tiepolo 11, I--34143 Trieste, Italy }
\affiliation{IFPU - Institut for fundamental physics of the Universe, Via Beirut 2, 34014 Trieste, Italy }
\affiliation{Scuola Normale Superiore, Piazza dei Cavalieri 7, I-56126 Pisa, Italy }

\author[0000-0003-2991-4618]{Nahum Arav}
\affiliation{Department of Physics, Virginia Tech, Blacksburg, VA 24061, USA}

\author[0000-0002-6748-2900]{Tiago Costa}
\affiliation{Max-Planck-Institut f\"ur Astrophysik, Karl-Schwarzschild-Stra\ss e 1, D-85748 Garching b. M\"unchen, Germany}

\author[0000-0002-9656-6281]{Kastytis Zubovas}
\affiliation{Center for Physical Sciences and Technology, Saul\'etekio al. 3, Vilnius LT-10257, Lithuania}
\affiliation{Astronomical Observatory, Vilnius University, Saul\'etekio al. 3, Vilnius LT-10257, Lithuania}

\author[0000-0003-2344-263X]{George Becker}
\affiliation{Department of Physics \& Astronomy, University of California, Riverside, CA 92521, USA}

\author[0000-0001-8582-7012]{Sarah E. I. Bosman}
\affiliation{Max-Planck-Institut für Astronomie, K\"onigstuhl 17, D-69117 Heidelberg, Germany}

\author[0000-0002-6830-9093]{Guido Cupani}
\affiliation{INAF - Osservatorio Astronomico di Trieste, Via G. B. Tiepolo 11, I--34143 Trieste, Italy}

\author[0000-0002-3324-4824]{Rebecca Davies}
\affiliation{Centre for Astrophysics and Supercomputing, Swinburne University of Technology, Hawthorn, Victoria 3122, Australia}
\affiliation{ARC Centre of Excellence for All Sky Astrophysics in 3 Dimensions (ASTRO 3D), Australia}

\author[0000-0003-2895-6218]{Anna-Christina Eilers}
\affiliation{MIT Kavli Institute for Astrophysics and Space Research, 77 Massachusetts Ave., Cambridge, MA 02139, USA}

\author[0000-0002-6822-2254]{Emanuele Paolo Farina}
\affiliation{Gemini Observatory, NSF's NOIRLab, 670 N A'ohoku Place, Hilo, Hawai'i, 96720, USA}

\author[0000-0002-9400-7312]{Andrea Ferrara}
\affiliation{Scuola Normale Superiore, Piazza dei Cavalieri 7, I-50126 Pisa, Italy}

\author[0000-0003-2754-9258]{Massimo Gaspari}
\affiliation{Department of Astrophysical Sciences, Princeton University, Princeton, NJ 08544, USA}

\author[0000-0002-5941-5214]{Chiara Mazzucchelli}
\affiliation{N\'ucleo de Astronom\'ia de la Facultad de Ingenier\'ia, Universidad Diego Portales, Av. Ej\'ercito Libertador 441, Santiago, Chile}

\author[0000-0003-2984-6803]{Masafusa Onoue}
\affiliation{Kavli Institute for Astronomy and Astrophysics, Peking University, Beijing 100871, China}
\affiliation{Kavli Institute for the Physics and Mathematics of the Universe (Kavli IPMU, WPI), The University of Tokyo, Chiba 277-8583, Japan}

\author[0000-0001-9095-2782]{Enrico Piconcelli}
\affiliation{INAF - Osservatorio Astronomico di Roma, via Frascati 33, 00040, Monte Porzio Catone, Italy}

\author[0000-0001-7883-496X]{Maria Vittoria Zanchettin}
\affiliation{SISSA, Via Bonomea 265, 34136, Trieste, Italy}
\affiliation{Dipartimento di Fisica, Sezione di Astronomia, Universit\'a di Trieste, via Tiepolo 11, 34143 Trieste, Italy}
\affiliation{INAF - Osservatorio Astronomico di Trieste, Via G. B. Tiepolo 11, I--34143 Trieste, Italy }

\author[0000-0003-3307-7525]{Yongda Zhu}
\affiliation{Department of Physics \& Astronomy,
    University of California, Riverside, CA 92521, USA}



\begin{abstract}
Luminous quasars are powerful targets to investigate the role of feedback from supermassive black-holes (BHs) in regulating the growth phases of BHs themselves and of their host galaxies, up to the highest redshifts. Here we investigate the cosmic evolution of the occurrence and kinematics of BH-driven outflows, as traced by broad absorption line (BAL) features, due to the C IV ionic transition. We exploit a sample of 1935 quasars at $z=2.1-6.6$ with bolometric luminosity log($L_{\rm bol}/$erg s$^{-1})\gtrsim46.5$, drawn from the Sloan Digital Sky Survey and from the X-shooter legacy survey of Quasars at Reionisation (XQR-30). We consider rest-frame optical bright quasars to minimise observational biases due to quasar selection criteria. We apply a homogeneous BAL identification analysis, based on employing composite template spectra to estimate the quasar intrinsic emission. We find a BAL quasar fraction close to 20\% at $z\sim2-4$, while it increases to almost 50\% at $z\sim6$. The velocity and width of the BAL features also increase at $z\gtrsim4.5$. We exclude that the redshift evolution of the BAL properties is due to differences in terms of quasar luminosity and accretion rate. These results suggest significant BH feedback occurring in the 1 Gyr old Universe, likely affecting the growth of BHs and, possibly, of their host galaxies, as supported by models of early BH and galaxy evolution.

\end{abstract}


\section{Introduction} \label{sec:intro}

Quasars are the brightest, non-transient sources in the Universe. They are powered by accretion onto super-massive black holes (BHs), whose emission typically dominates over the host-galaxy emission in the rest-frame UV and optical bands. High-redshift quasars at $z\simeq2-6.5$ provide a unique window on the growth phases during which the most massive BHs and their host-galaxies assembled the bulk of their mass \citep[][]{Marconi04, Volonteri06}.
Luminous quasars at $z\gtrsim2$, with bolometric luminosity $L_{\rm Bol}\sim10^{47}$ erg/s, are typically powered by BHs with masses of $10^8-10^{10}$ M$_\odot$ yr$^{-1}$ and high accretion rates $\lambda_{\rm Edd}=L_{\rm Bol}/L_{\rm Edd}\sim0.01-1$ \citep[e.g.][]{Kurk07,Shen11,Mazzucchelli17}, where $L_{\rm Edd}$ is the Eddington luminosity. The large radiative output of these BHs is expected to drive powerful outflows, able to regulate the BH and host-galaxy growth, by injecting large amounts of energy and momentum in the galaxy interstellar medium \citep[ISM,][]{Zubovas&King12,Gaspari17, Menci19}.

BH-driven outflows in high-redshift quasars are often revealed by blueshifts of high-ionization emission lines with respect to the quasar systemic redshift \citep{Shen19, Meyer19, Schindler20} and from broad/asymmetric wings in the emission line profiles \citep[e.g.][]{Zakamska16,Kakkad20}. Another powerful tracer of BH-driven outflows are broad ($>2,000$ \kms) absorption line (BAL) systems occurring in the rest-frame UV spectrum \citep{Weymann91}, blueward of high-ionisation emission lines such as C IV $\lambda1,549$\AA, Si IV $\lambda1,397$\AA, N V $\lambda1,240$\AA\ (HiBAL),  and of low-ionisation emission lines such as Mg II $\lambda2,800$\AA (LoBAL). BAL features have been indeed identified in quasars at all redshifts up to $z>7$ \citep{Wang18,Wang21}. Observational studies, mostly based on spectra from the Sloan Digital Sky Survey (SDSS), found that BAL outflows are observed in 10-17\% of $z\simeq2-4$ quasars, and have typical velocities of 5,000-10,000 km/s \citep{Gibson09,Paris18}, although a few percent of them can reach $10-15$\% of the light speed \citep{Hamann18,Bruni19,Rodriguez-Hidalgo20}. LoBAL outflows are typically observed in a small fraction ($10-15$ \%) of BAL quasars.

In luminous quasars at $z\simeq2-3$, BAL outflows have been found to carry kinetic power values in the range 0.1-10\% of $L_{\rm Bol}$ \citep{Dunn10,Borguet13,Byun22}, consistent with expectations for an efficient BH-feedback mechanism \citep{Fiore17}. Similarly, galaxy evolution models identify BAL outflows as an important source of feedback \citep[e.g.][]{Costa14,Torrey20}.
Different scenarios for the launching and geometry of BAL outflows have been proposed \citep{Elvis00,Proga00,Proga&Kallman04, Xu20,Zeilig-Hess20}. Although these winds are expected to originate within a pc from the BH \citep{Sadowski17}, close to the accretion disc, observations indicate that BAL absorption in luminous quasars often occurs at much larger distances from the BH, that is 100-1,000 pc \citep{Arav18,Hemler19}. This suggests that these ionised outflows represent an important source of BH feedback on the growth of the BH itself and of the host-galaxy.

At very high redshift ($z\gtrsim6$), the occurrence and properties of BAL quasar outflows are much less explored. The number of discovered BAL quasars has kept increasing with the growing number of quasars with spectroscopic observations of sufficient quality to probe at least the most prominent BAL features. A first guess for a BAL quasar fraction as high as 50\% at $z\sim6$ had been previously suggested by \cite{Maiolino04}, based on a very small sample of eight quasars. A BAL fraction of 16-24 \% was later reported by \cite{Shen19,Schindler20,Yang21}, although the BAL identification was mostly based on visual inspection and/or limited spectral range and SNR of the quasar spectra. In \cite{Bischetti22}, we performed the first systematic investigation of BAL outflows in 30 quasars at $5.8\le z\le 6.6$ from the X-shooter legacy survey of Quasars at the Reionisation epoch \citep[\href{https://xqr30.inaf.it/}{XQR-30}, ][]{Zhu21,Zhu22,Bosman22,Lai22,Chen22}, finding that about 47\% of XQR-30 quasars show BAL features associated with C IV, a fraction that is $\sim$2.4 times higher than what observed in $z\sim2-4$ quasars. The majority of BAL outflows at $z\sim6$ also show extreme velocities of 20,000-50,000 km s$^{-1}$, rarely observed at lower redshift. These findings are indication of an evolution in BAL properties between $z\simeq2$ and $z\simeq6$ quasars.

However, different factors such as $L_{\rm Bol}$ or $\lambda_{\rm Edd}$ are believed to affect the fraction of BAL quasars and their kinematics  \citep[e.g.][]{Dai08, Allen11, Bruni19}.
In this paper we investigate the BAL fraction and the BAL kinematics dependence on redshift and on quasar nuclear properties (\lbol, $\lambda_{\rm Edd}$), with the goal of identifying the main driver of the varying BAL properties. To this purpose, we measure the occurrence and kinematic properties of BAL outflows associated with the C IV ionic species in a sample of luminous quasars in the redshift range $z=2.1-6.6$, with matched selection criteria, and by adopting a homogeneous BAL identification analysis. 

The paper is structured as follows. In Sect. \ref{sect:sample}, we describe the quasar sample, including quasar selection criteria (Sect. \ref{subsec:selection}). Section \ref{sect:ident} details the BAL identification analysis and compares it with previous approaches. The main results on the BAL fraction and on the BAL kinematics are presented in Sect. \ref{sec:balfrac} and Sect. \ref{sect:kine}, respectively. We discuss the cosmic evolution of the BAL properties and its implications for BH and galaxy evolution in Sect. \ref{sect:discussion}. A summary and the main conclusions are given in Sect. \ref{sec:conclusion}.
Throughout this paper, we adopt a $\rm \Lambda CDM$ cosmology with $H_0=67.3$ km s$^{-1}$, $\rm\Omega_\Lambda=0.69$ and $\rm\Omega_M = 0.31$ \citep{Planck16}.

\begin{deluxetable}{lcc}\label{table:literature}
\tablenum{1}
\tablecaption{BAL fraction estimates at $2<z<4$}
\tablewidth{0pt}
\tablehead{
\colhead{Reference} &  \fobs\  &  \fint\  
}
\decimalcolnumbers
\startdata
\multicolumn{3}{c}{UV selected quasars}\\
    \citet{Reichard03}   & 14.0$\pm1.0$\% & 13.4$\pm$1.2\%  \\
    \citet{Knigge08}   & 13.7$\pm0.3$\% & $\le23$\%  \\
    \citet{Gibson09}   & 15.1$\pm$0.6\% & 18.5$\pm$0.7\% \\
    \citet{Allen11}    & 8.0$\pm$0.1\% & 40.7$\pm5.4$\% \\
    \citet{Hewett03}    & 15$\pm$3\% & 22$\pm$4\% \\
    \hline
    \multicolumn{3}{c}{Optically selected quasars}\\
    \citet{Dai08}      &  $\sim20$\%  & 23$\pm$3\%\\
    \cite{Ganguly08}   &     23\%       & $-$ \\
    \cite{Maddox08}      &   17.5     & 18.5\% \\ 
\enddata
\tablecomments{(1) Reference, (2) observed BAL fraction, (3) intrinsic BAL fraction (see Sect. \ref{subsec:selection}). }
\end{deluxetable}

\section{Quasar sample}\label{sect:sample}

\subsection{Quasar selection}\label{subsec:selection}
The observed fraction of BAL with respect to non-BAL quasars is the convolution of the intrinsic BAL quasar occurrence and a quasar selection function. Standard quasar selection algorithms in optical surveys rely on multicolor imaging data which probe the rest-frame UV quasar spectrum at $z>2$ \citep[e.g.][]{Richards02}. These UV colors can be affected by
BAL troughs associated with high-ionisation species such as C IV. In addition, BAL quasars typically show redder UV continua than non-BAL quasars \citep[e.g.][]{Reichard03,Gibson09}. Depending on redshift, BAL quasars may thus show UV colors resembling those of stars, and may be more easily missed by  spectroscopic samples than non-BAL quasars.

Aiming to measure the intrinsic fraction of BAL quasars across cosmic time, two main approaches can be followed. One possibility is to translate the observed BAL fraction (\fobs) into an intrinsic fraction (\fint) by correcting  for  selection effects. Previous attempts to correct for these effects, however, resulted in very different \fint\ (by a factor of three), and different \fobs\ to \fint\ corrections (by a factor of five), even considering the same set of selection effects, and similar redshift intervals. When considering quasars from SDSS and from the Large Bright Quasar Survey, \fint(\fobs) values in the range 13.4(8.0)\% to 40.7(23\%) have been reported, as summarised in Table \ref{table:literature}.
In all these cases, the derived \fint\ mostly depends on the adopted assumptions  on the input BAL properties, such as the distributions of the velocity, width and depth of the absorption, and on the relation between the above quantities and reddening. The spread of plausible assumptions naturally leads to large uncertainties. The true \fint\ could in principle be derived by quantifying the selection effects on a set of synthetic spectra based on a physical model of BAL quasar outflows. Such an approach would nonetheless suffer from large uncertainties, due to the lack of a self-consistent physical model, able to widely reproduce the properties of the BAL quasar population \citep[e.g.][]{Elvis00,Xu20}.

In this work, we follow a different approach that allows us to bypass the above issues. As we are mainly interested in the redshift evolution of the BAL fraction and kinematic properties, we apply a similar selection function at all redshifts. In particular, we consider rest-frame optically bright quasars to measure the BAL fraction in the redshift interval $z\simeq2-6.5$. At $z\simeq2-4$, the rest-frame optical band is probed by near-infrared surveys such as 2MASS \citep{Skrutsie06} and UKIDSS \citep{Warren07} in the H and K bands. For higher redshift quasars ($z\sim6$), a similar spectral coverage is provided by WISE \citep{Wright10} in the W1 and W2 bands \citep[see][for details]{Bischetti22}.
The rest-frame optical selection allows us to minimise selection effects by avoiding those biases generated by UV absorption troughs in the colour selection. The \fobs-to-\fint\ correction in rest-frame optical selected quasars is indeed expected to be modest (10-20\%): \cite{Dai08} found a \fint=23±3\% in SDSS and 2MASS, K-band bright quasars at $z\simeq2-4$ (\fobs$\sim20$\%). This fraction is similar to that found by \cite{Ganguly07} and \cite{Ganguly08} when combining SDSS and 2MASS catalogs (23\%). Again, \cite{Maddox08} reported a \fint = 18.5\% in SDSS and UKIDSS bright quasars, considering the fraction of quasars missed by the UKIDSS survey (\fobs = 17.5\%). As a general trend, the observed BAL fraction calculated in rest-frame optical selected quasars is roughly 50\% higher than the BAL fraction observed in quasars selected only in the rest-frame UV: $f_{\rm obs}^{\rm optical}\simeq1.5 \times f_{\rm obs}^{\rm UV}$. In addition, the observed and intrinsic BAL fraction calculated in rest-frame optical selected quasars are similar. They are also similar to the intrinsic BAL fraction calculated in quasars selected only in the rest-frame UV: $f_{\rm obs}^{\rm optical}\simeq f_{\rm int}^{\rm optical} \sim f_{\rm int}^{\rm UV}$.

The \fobs\ of quasar samples selected from X-ray to radio surveys, affected by a variety of different selection effects, is typically lower or similar to 20\% \citep[e.g.][]{Giustini08,Becker00},  supporting the fact that an optical rest-frame selection limits selection biases against BAL quasars. In general, BAL fractions higher than 20\% have been reported only
i) when considering the absorption index \citep[AI,][]{Hall02} criterion to identify BAL quasars, instead of the balnicity index criterion considered in this work \citep[BI,][see Sect. \ref{subsec:balprop}]{Weymann91}. The AI criterion identifies as BAL troughs the absorption features that are wider than 1,000 km/s, and  typically results in a two times higher BAL fraction than using $BI>0$ \citep{Dai08}. Although the AI criterion can be used to identify the shallowest absorption features, it results in a large fraction of false BAL identifications when applied to spectroscopic data with modest spectral resolution \citep{Knigge08}.
ii) in the reddest quasars \citep{Urrutia08,Glikman12}, which however constitute a minority ($\sim$10\%) of the rest-frame optical selected quasars.

Given the above evidence, the BAL fractions presented in this work and their redshift evolution are robust against selection effects.

\begin{figure}[thb]
    \centering
    \includegraphics[width = 1\columnwidth]{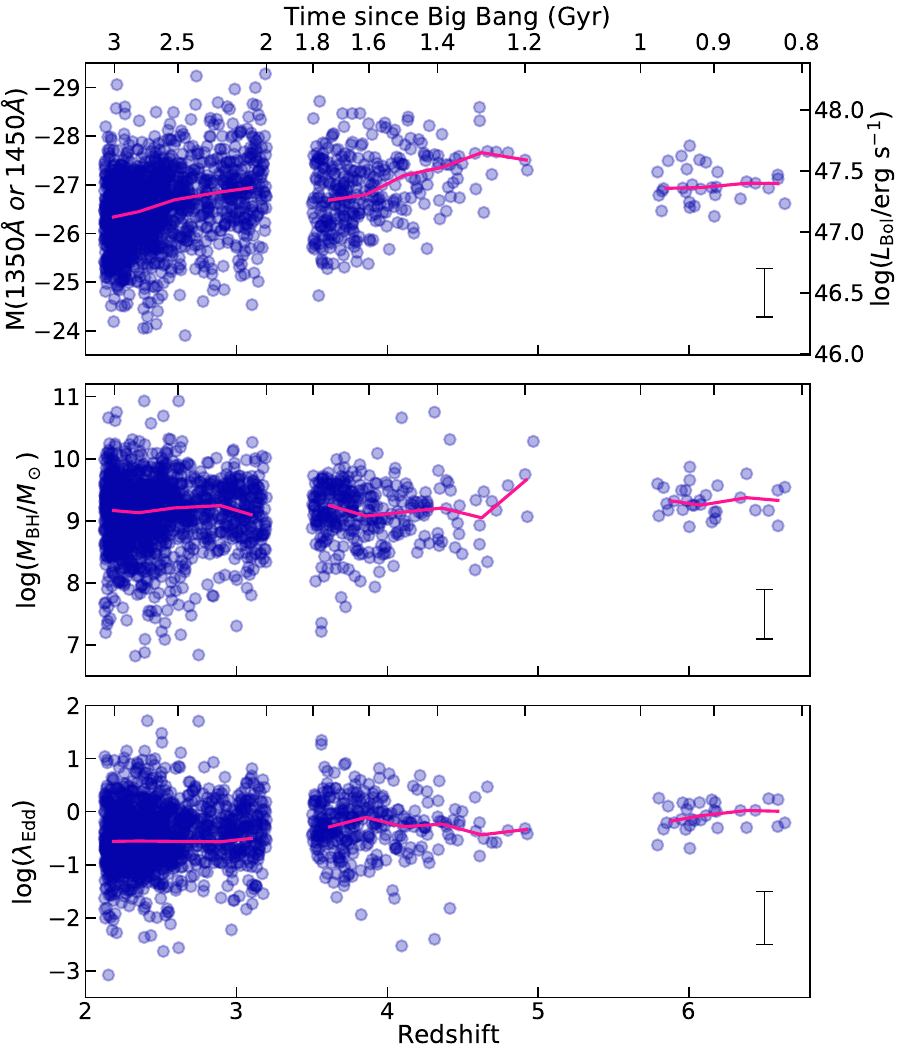}
    \caption{Absolute magnitude (top), BH mass (middle) and Eddington accretion rate (bottom) as a function of redshift for the quasar sample considered in this work. Top axis indicates time since Big Bang. Magenta line represents the median magnitude in redshift bins $\Delta z$= 0.25.  Error bars represent the typical uncertainties (see Sect. \ref{subsec:low-z} and Sect. \ref{subsec:high-z})}.
    \label{fig:M-z}
\end{figure}

\subsection{$2.1<z<5.0$ sample}\label{subsec:low-z}
The low-redshift sample consists of 1578 quasars at $2.13<z<3.20$ from the catalogue of SDSS DR7 quasars by \cite{Shen11}, selected to be detected by 2MASS in the H(1.7$\mu$m) and K(2.2$\mu$m) bands, which cover rest-frame optical regions of the quasar spectrum, similar to those probed by W1 and W2 in the high-redshift sample.  
We also include 327 SDSS DR7 quasars at $3.6<z<5.0$ from the \cite{Shen11} catalogue, requiring them to be detected in the 2MASS K band (which correspond to W1 at $z\sim6$). These quasars are about a factor of ten more luminous than the average SDSS quasar at $z>2.13$, with a median magnitude of $M_{1450\rm \AA}=-26.6$  ($-26.5$ at $z<3.2$, $-26.9$ at $3.6<z<5$, Fig. \ref{fig:M-z}). Their spectra have a typical SNR $\sim18$ per 70 \kms\ pixel in the $1500-1600$ \AA\ range. We derive \lbol\ by applying a bolometric correction to the monochromatic luminosity of the rest-frame UV continuum \citep[Fig. \ref{fig:M-z}, with a typical scatter of $\sim0.2$ dex,][]{Richards06,Runnoe12}. This sample benefits from measured black hole masses \citep{Shen11}, which are based on the Mg II line for $z\lesssim2.3$ quasars (349 quasars) and on the C IV emission line (1160 quasars) at $z>2.3$, including the empirical correction for non-virial motions by \cite{Coatman17}. We verified that MgII and C IV BH masses are consistent, for the 340 quasars in which both tracers are available, within the respective $\sim0.3-0.5$ dex uncertainties \citep[e.g.][]{Vestergaard&Osmer09},  in agreement with the results of \cite{Shen11}.
For the subsample of quasars already discovered by SDSS DR5, the occurrence and properties of BAL outflows were investigated by \cite{Gibson09}, providing us with a reference analysis to test our BAL identification method. For the remaining quasars, BAL identification in the  catalogue was based on visual inspection \citep{Shen11}. In this work and in \cite{Bischetti22}, we have re-analysed all SDSS quasars in the $2.1<z<5.0$ sample, following the approach described in Sect. \ref{sect:ident}.

\subsection{$5.8<z<6.6$ sample}\label{subsec:high-z}
The high-redshift sample used in this work consists of 30 luminous quasars (median AB magnitude $M_{1450\rm\AA}=-26.9$) at $5.8\lesssim z\lesssim 6.6$ from the XQR-30 survey. Their magnitude and redshift distributions are shown in Fig. \ref{fig:M-z} \citep{Banados16,Mazzucchelli17,Reed17,Decarli18,WangF19,Eilers20}.
These quasars are selected to be bright in both the rest-frame UV (AB magnitude $J\le19.8$ for $z<6.0$ sources and $J\le20.0$ for quasars at $z\ge6.0$) and in the rest-frame optical, being all detected by WISE in the W1(3.4$\mu$m) and W2(4.6$\mu$m) bands \citep{Banados16, Ross20}. Further details about the selection of \xqr\ quasars, data reduction of the \xshoot\ spectra, and the target list are given in \cite{Bischetti22}. This sample benefits from deep \xshoot\ observations (SNR $\gtrsim$ 25 per 50 \kms\ pixel in the rest-frame spectral range 1600-1700 \AA), and  robust, Mg II-based black hole masses \citep{Lai22}. As for the low-redshift sample, we verified the consistency between Mg II and C IV-based black hole masses. Bolometric luminosities are computed via rest-frame UV bolometric correction using the same method applied to the low-redshift sample \citep{Runnoe12}.

A systematic search and characterisation of C IV BAL outflows in XQR-30 was performed in \cite{Bischetti22}, based on the same approach used in this work (Sect. \ref{sect:ident}).
Although \xshoot\ spectra of similar quality (SNR $\gtrsim$ 10) exist to date for several luminous quasars at $5.0<z<5.4$, we do not consider them in this work as they have been mostly selected for studies of intervening absorbers and are therefore biased against the presence of BAL features \citep{Becker19}.

\begin{figure}[thb]
    \centering
    \includegraphics[width=1\columnwidth]{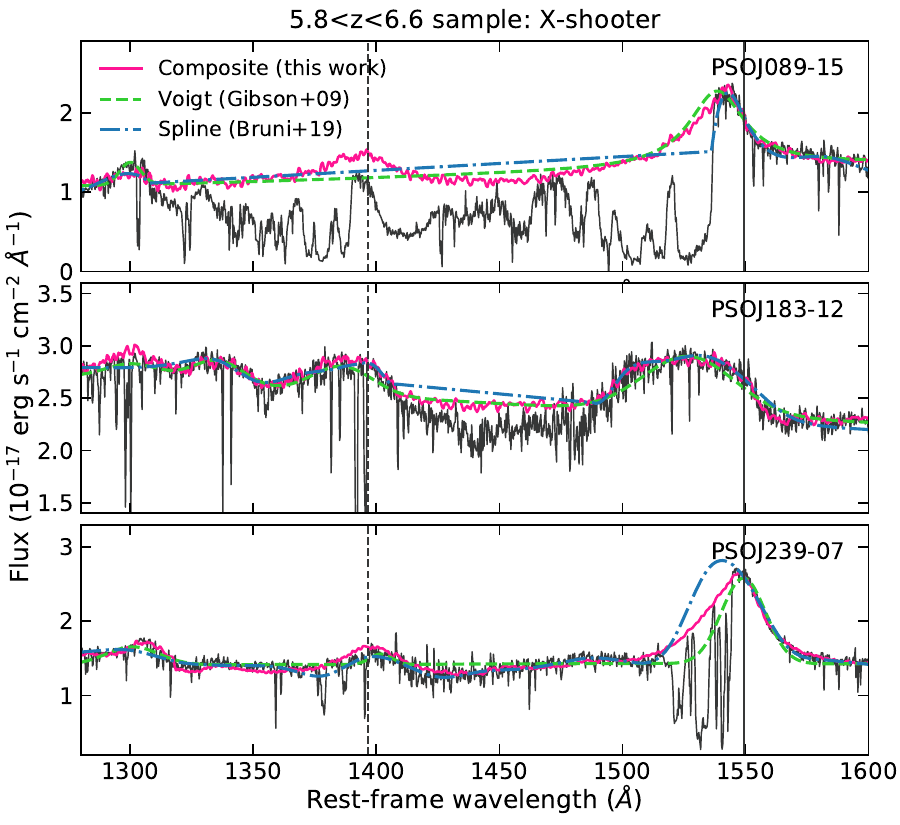}
    \caption{Example of three BAL quasars from the $5.8<z<6.6$ sample considered in this work. We show the comparison between our estimate of the intrinsic quasar emission, based on composite template spectra (Sect. \ref{subsec:comp}), with the quasar emission model by \cite{Gibson09} and the spline model by \cite{Bruni19}. Vertical solid and dashed lines refer to the rest-frame wavelength of C IV and Si IV, respectively.}
    \label{fig:comparison}
\end{figure}

\begin{figure*}[thb]
    \centering
    \includegraphics[width=1\textwidth]{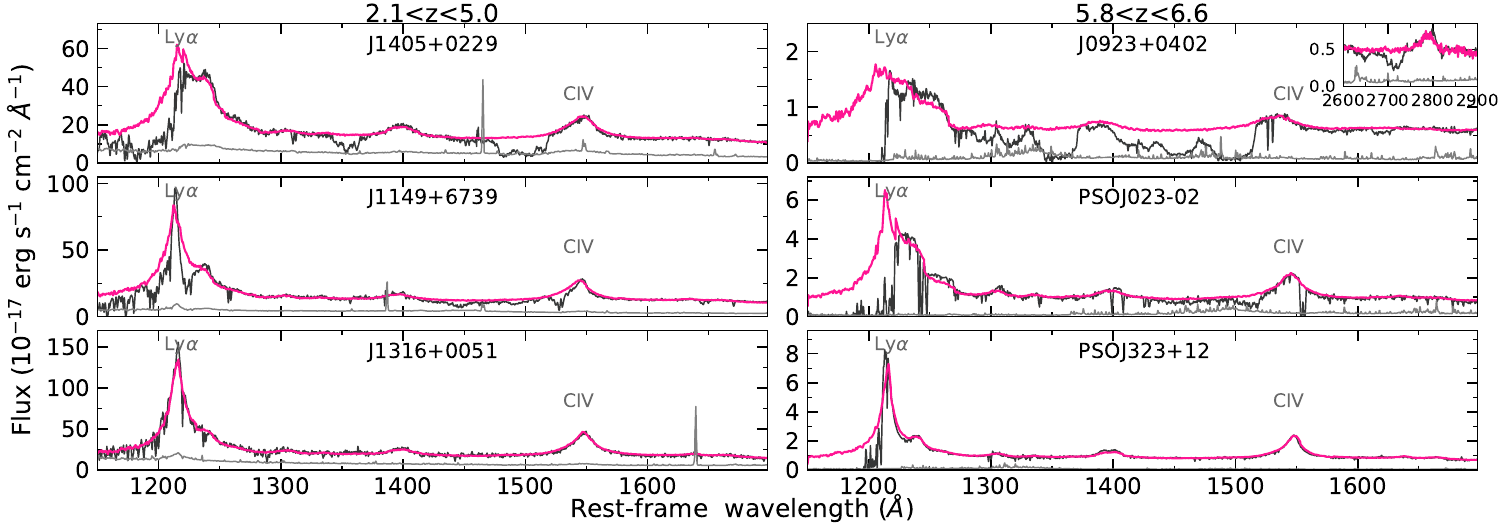}
    \caption{Examples of BAL (top and middle) and non-BAL (bottom) quasar spectra, randomly selected from our sample and binned to 50 \kms\ (black curves). Left(right) column shows SDSS(X-shooter) spectra. In the case of J0923+0402 ($z=6.626$), the inset displays the spectral region close to Mg II affected by BAL features. Composite templates, used to estimate the intrinsic quasar emission, are shown by the magenta curve (Sect. \ref{subsec:comp}).  We also show in grey the flux uncertainty multiplied by a factor of five}. Labels indicate the Ly$\alpha$ and  C IV emission lines. 
    \label{fig:composite1}
\end{figure*}

\section{Identification of BAL quasars}\label{sect:ident}
\subsection{Reconstruction of quasar continuum emission}\label{subsec:comp}

To search for BAL absorption troughs, the first step is to model the intrinsic continuum emission in the rest-frame UV. A possible approach is to fit the spectra with a quasar emission model accounting for continuum and line emission, and extrapolating this model to the spectral region affected by the BAL (Fig. \ref{fig:comparison}).  Models typically include a power-law component to reproduce the quasar continuum emission, and Voigt or gaussian profiles to account for emission lines \citep[][]{Gibson09, Shen11, Shen19}. Alternatively, more complex, non-physically motivated functions such as splines are sometimes used, as they can be more flexible in reproducing a variety of quasar spectra \citep[e.g.][]{Bruni19}. Both approaches require that a sufficiently large portion of the quasar spectrum is free from both absorption and emission lines, allowing to anchor the fit. However, this condition is often unsatisfied in the spectral region between Ly-$\alpha$ and C IV, resulting in large uncertainties (Fig. \ref{fig:comparison}). In addition, both methods are largely dependent on the regions of the spectrum that are masked prior to fitting, especially in those cases where BAL features cover a wide spectral range or affect emission line profiles.
 
These uncertainties can be minimised using composite templates built from observed quasar spectra. Quasar spectra can be combined via different techniques such as principal component analysis \citep{Trump06,Paris18,Guo19} or non-negative matrix factorisation \citep{Allen11}. Here we build, for each quasar of the sample, a composite template based on a large number of SDSS quasar spectra classified as non-BAL in \cite{Gibson09}, each matching within $\pm$ 20\% its color and the equivalent width (EW) of the C IV line \citep{Shen11}. Given the  anti-correlation between the EW of C IV and its blueshift with respect to the quasar redshift due to outflowing gas motions \citep{Coatman17,Vietri18}, the latter criterion allows us to reproduce all levels of asymmetry in the C IV line profile.  A similar approach was recently adopted by \cite{Wang18,Wang21}, who created composite templates based on matching C~IV blueshift of two $z>7$ BAL quasars. 
The C~IV EW or the blueshift criteria should in principle produce similar composite templates. However, SDSS quasar redshifts rely on automatic fitting procedures based on a limited spectral coverage, and are therefore affected by uncertainties on the C IV blueshift that can reach thousands of \kms, leading to an inaccurate selection of the quasar spectra contributing to the template. We find that a better spectrum-template agreement can be achieved by using the C IV EW during the selection of the quasar spectra.

Concerning the quasar colours, we consider the F(1700\AA)/F(2100\AA) flux ratio to reproduce the continuum slope redwards of C IV, and the F(1290\AA)/F(1700\AA) flux ratio, to account for a change in the continuum slope as observed in red quasar spectra \citep[e.g.][]{Trump06,Shen19}. We calculate F(1700\AA) and F(2100\AA) as median flux values over 100 \AA\ and F(1290\AA) over 30 \AA\ in the rest-frame. The underlying assumption is that the observed and intrinsic continuum emission do not differ around 1290 \AA\, that is the bluest spectral region, redward of Ly-$\alpha$, free from strong emission lines. In fact, Ly-$\alpha$ forest absorption complicates our measure of the $\lambda<1216$ \AA\ quasar continuum, owing to the absorption becoming stronger with increasing redshift. In the case that BAL absorption troughs affect the spectral region close to 1290 \AA, our approach would provide a lower limit on the intrinsic quasar continuum emission and on the absorption $BI$ (Sect. \ref{sect:sample}).

\begin{figure}[htb]
    \centering
    \includegraphics[width=1\columnwidth]{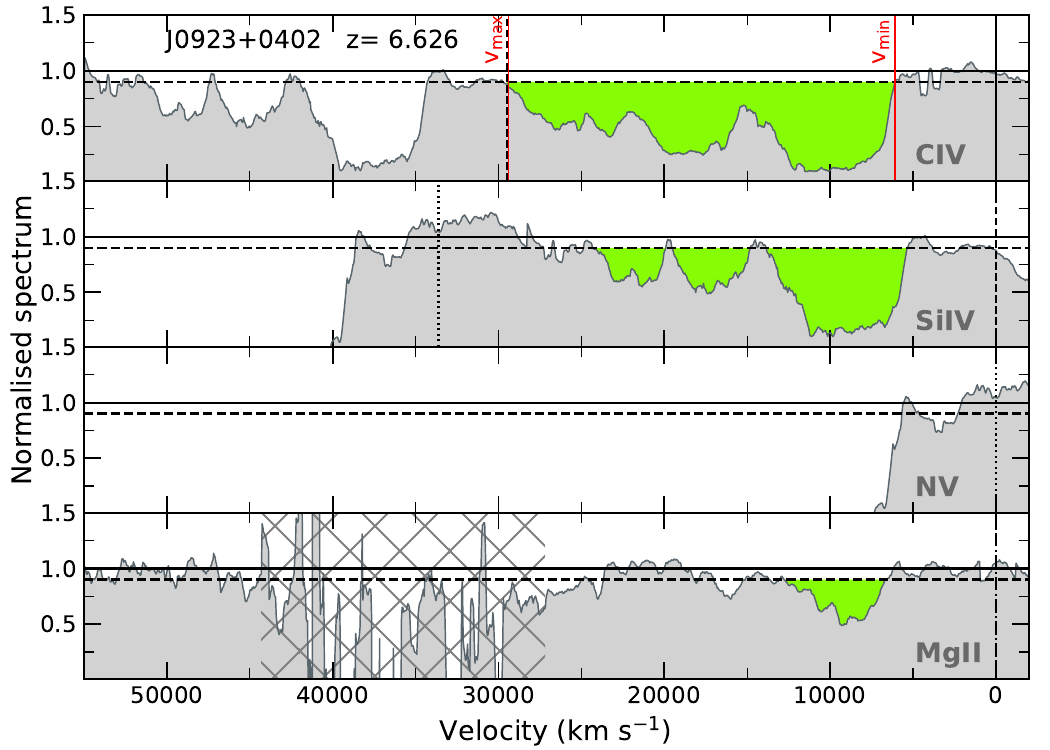}
     \includegraphics[width=1\columnwidth]{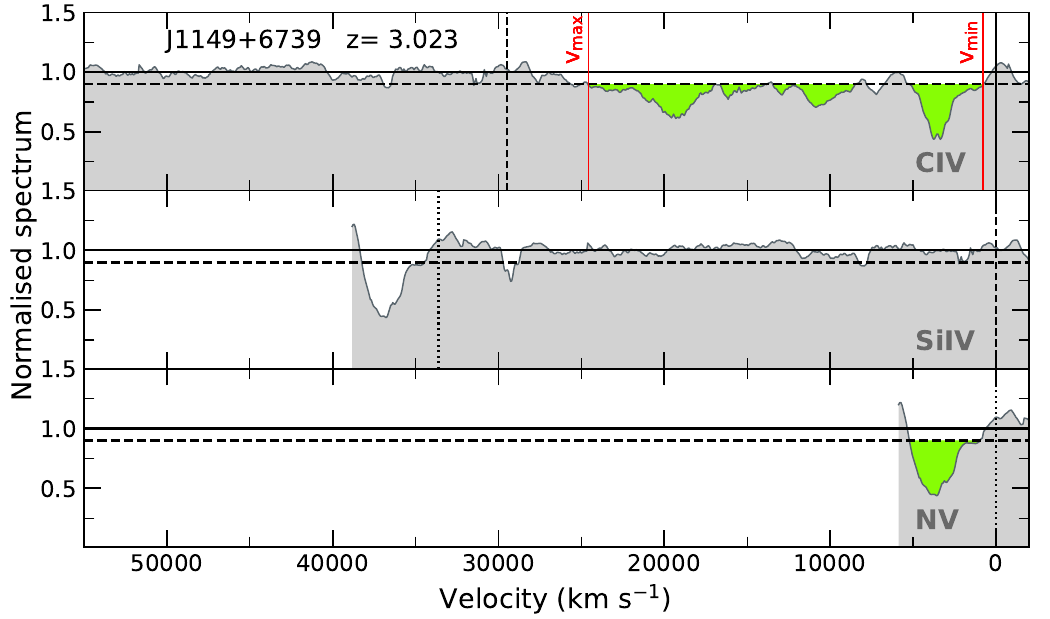}
     \includegraphics[width=1\columnwidth]{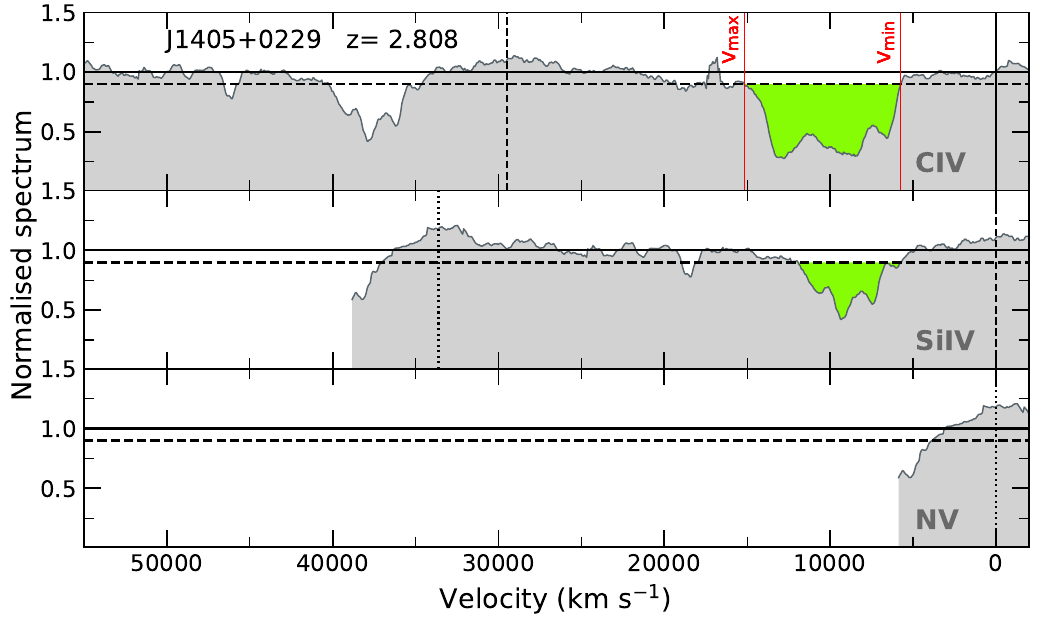}
     \caption{Examples of normalised spectra for three BAL quasars in our sample, smoothed to 500 \kms. The velocity axis in each panel is relative to the rest-frame wavelength of the ionic species indicated by the label. Vertical solid, dashed, dotted and dashed-dotted lines indicate the velocity associated with C IV, Si IV, N V and Mg II emission lines, respectively. The solid(dashed) horizontal line represents a flux level of 1.0(0.9). BAL troughs, corresponding to a flux level $<0.9$ (Eq. \ref{eq:bi}), are highlighted as green shaded areas. Vertical red lines indicate \vmax\ and \vmin\ for the C IV BAL. The hatched area indicates the spectral region affected by strong telluric features.} 
    \label{fig:norm}
\end{figure}

Starting from the total \cite{Shen11} catalogue of 11800 SDSS quasars in the redshift range $2.13~<~z~<~3.20$, for which SDSS spectroscopy probes the $1216-2100$ \AA\ wavelength range, the composite template spectrum is built as the median of a hundred randomly selected, non-BAL \citep{Gibson09,Shen11} quasar spectra. The above number is a trade-off between using stringent selection criteria and ensuring that the composite template is not affected by individual quasar spectra. The composite template is normalised to the median flux value of the quasar spectrum  in the rest-frame  1650-1750 \AA\ spectral interval, avoiding prominent emission lines and strong telluric absorption for the redshift interval covered by our sample \citep{Smette15}. Figure \ref{fig:comparison} shows that the above approach provides us with (i) a solid reconstruction of the C IV profile even when affected by strong absorption features, (ii) a reasonable estimate of the quasar emission blueward of C IV also in the most absorbed spectra, (iii) a  conservative estimate of the continuum emission, typically lower or similar to the continuum  reproduced by the \cite{Gibson09}, \cite{Bruni19} models. Figure \ref{fig:composite1} shows examples of the SDSS and \xshoot\ spectra for the quasars belonging to the $2.1<z<5.0$ and $5.8<z<6.6$ samples, respectively. The \xshoot\ spectra and composite templates for the remaining $5.8<z<6.6$ quasars are shown in Fig. \ref{fig:composite-appx}. Normalised spectra are obtained by dividing each spectrum by its matched composite template (Fig. \ref{fig:norm}).


\subsection{Characterisation of BAL troughs}\label{subsec:balprop}
The standard indicator used to identify absorption features in quasar spectra due to BAL outflows is the balnicity index (BI), firstly introduced by \cite{Weymann91}, which is a modified equivalent width of the BAL absorption and is less affected by false BAL identification  than the AI criterion \citep{Knigge08}. Here we adopt the balnicity index definition by \citep{Gibson09}:
\begin{equation}\label{eq:bi}
    BI=\int_0^{v_{\rm lim}}\left(1-\frac{f(v)}{0.9}\right)Cdv    
\end{equation}
where $f(v)$ is the normalised spectrum, $C=1$ if $f(v)<0.9$ for contiguous troughs of $>2,000$ \kms, $C=0$ otherwise. The $f(v)<0.9$ threshold identifies as BAL features only the spectral regions dipping by 10\% or more below the quasar continuum level, and represents a trade-off between taking into account typical uncertainties on the level of the modelled quasar continuum and including shallow absorption features \citep{Weymann91}.

We search the normalised spectra for BAL features in the region between Ly-$\alpha$ and C IV, that is corresponding to  \vlim$\sim64,000$ \kms\ for C IV BAL outflows, $\sim35,000$ \kms\ for Si IV and $\sim6,000$ \kms\ for N V, respectively. We do not extend the search to smaller wavelengths due to the stronger intergalactic medium absorption blueward of Ly-$\alpha$ with increasing redshift, preventing us from a homogeneous BAL identification at different epochs. We identify the minimum (\vmin) and maximum (\vmax) BAL velocity for a given transition, as the lowest and highest velocity for which $C=1$ in Eq. \ref{eq:bi}, respectively. 
Previous studies were typically limited to $v_{\rm lim}=25,000$ \kms\ \citep{Trump06,Gibson09,Guo19}, to avoid discriminating between BAL outflows associated with C IV and Si IV ions. Here we exploit the fact that the C IV optical depth is usually similar or larger than the Si IV depth in BAL quasars \citep{Dunn12}. This allows us to use the velocity range of the C IV BAL troughs to identify absorption associated with Si IV  \citep[Fig. \ref{fig:norm}, see also][]{Wang18,Bruni19}. This implies that absorption features blueward of Si IV are due to a low velocity Si IV BAL if a C IV BAL with similar velocity is observed. Otherwise, these features are considered to be produced by a high-velocity ($v_{\rm max}\gtrsim30,000$ \kms) C IV BAL \citep[see also][]{Rodriguez-Hidalgo20}. We similarly assume that the C IV optical depth is larger than those of Mg II and N V and apply the methodology above to identify Mg II and BAL features. We measure the BAL width, defined as \wmax\ = \vmax $-$ \vmin. The distributions of \bi, \vmin\, and \vmax\ for the C IV BAL quasars identified in our sample are shown in Fig. 2 of \cite{Bischetti22}, and values for individual quasars at $5.8<z<6.6$ are listed in their Table E1. Uncertainties on the BAL parameters have been computed taking into account the uncertainty on both slope and normalisation of the best-fit composite template, following the method in \citep{Bischetti22}. Briefly, we created for each quasar a bluer and a redder template as the  median of the 33\% bluest and the 33\% reddest spectra contributing to the best-fit template, and we used them to create new normalised spectra, from which the range of variation of the BAL parameters is calculated. The median uncertainties (68\% confidence level) on \bi, \vmin\, \vmax, and \wmax\ are 310, 200, 380, and 340 \kms, respectively, while uncertainties for individual quasars in the $5.8<z<6.6$ sample are given in Table \ref{table:otherbal}.  We note that adopting a lower threshold of $\sim0.8$ in Eq. \ref{eq:bi} would identify as BAL only the deepest absorption features (corresponding to a \bi$\gtrsim1000$ \kms), and would typically imply lower \vmax(higher \vmin) by $\sim1000-2000$ \kms. 
\begin{deluxetable}{lcccc}
\tablenum{2}
\tablecaption{Si IV and N V BAL parameters}
\label{table:otherbal}
\tablewidth{0pt}
\tablehead{
\colhead{Name} &  Species & \bi\ & \vmin\ &  \vmax\  
}
\decimalcolnumbers
\startdata
        PSOJ009$-$10    & Si IV & 2,880$^{+750}_{-160}$ & 33,040$^{+220(*)}_{-110}$ & 38,360$^{+150(*)}_{-1360}$ \\
        PSOJ065$+$01   & Si IV   & 660$^{+430}_{-150}$ & 17,240$^{+130}_{-210}$ & 27,190$^{+1030}_{-510}$ \\
        \multirow{2}{*}{PSOJ089$-$15}    & Si IV & 8,150$^{+710}_{-620}$ & 2,280$^{+150}_{-120}$ & 28,040$^{+160}_{-220}$  \\ 
                        & N V & 2,280$^{+190}_{-280}$ & 2,270$_{-110}^{+150}$ & 5,870$_{-170}^{(**)}$ \\
        J0923$+$0402    & Si IV  & 7,970$^{+230}_{-150}$ & 6,110$^{+150}_{-150}$ & 24,330$^{+350}_{-200}$ \\ 
        PSOJ217$-$07    & Si IV & 440$^{+170}_{-150}$ & 30,460$^{+270}_{-120}$ & 35,000$^{+430}_{-100}$ \\ 
        PSOJ231$-$20   & N V & 630$^{+230}_{-190}$ & 140$^{+130}_{-210}$ & 2,350$^{+160}_{-140}$ \\
        \multirow{2}{*}{PSOJ239$-$07}    & Si IV & 230$^{+310}_{-120}$ & 830$^{+160}_{-120}$ & 3,360$^{+1010}_{-120}$ \\          
                        & N V  & 1,810$^{+130}_{-100}$ & 820$^{+100}_{-120}$ & 5,720$^{+110}_{-100}$ \\
        J2211$-$3206    & Si IV & 3,420$^{+130}_{-110}$ & 9,550$^{+290}_{-470}$ & 19,120$^{+350}_{-140}$ \\        
        J2250$-$5015   & Si IV & 1,320$^{+170}_{-110}$ & 29,460$^{+150}_{-810}$ & 37,540$^{+170}_{-130}$\\
\enddata
\tablecomments{1) quasar ID, (2) Ionic species of the BAL, (3-5) balnicity index, minimum and maximum velocity  of the Si IV BAL outflows, in units of \kms. Positive \vmin\ and \vmax\ values indicate blue-shifted absorption. $^{(*)}$ We assume the same range of \vmin\ and \vmax\ of the C IV BAL \citep{Bischetti22}.  $^{(**)}$ \vmax\ corresponds to rest-frame $1216$ \AA, below which the spectrum is dominated by Ly-$\alpha$ forest absorption (Sect. \ref{sect:ident})}.
\end{deluxetable}

\subsection{Non-C IV BALs in the XQR-30 sample}
Although the identification of BAL features in high-redshift quasars is mostly based on C IV, due to its high optical depth, the combined study of several other ionic species can be used to investigate the ionisation level, kinematic structure and column density of the outflow \citep[e.g.][]{Feliz14,Baskin15}. However, only a few cases of non-CIV BAL outflows at $z\gtrsim6$ have been reported so far \citep{Wang18, Wang21}.

Here we find that, of the 14 C IV BAL quasars identified in \citep{Bischetti22}, eight also show a BAL system associated with the Si IV ion (namely PSOJ009-10, PSOJ065+01, PSOJ089-15, J0923+0402, PSOJ217-07, PSOJ239-07, J2211-3206, J2250-5015), three show a N V BAL (namely PSOJ089-15, PSOJ231-20, PSOJ239-07), and one (J0923+0402) shows a Mg II BAL. Figure \ref{fig:norm} shows as an example the normalised spectrum of quasar J0923+0402 at $z\sim6.6$, for which we identify strong BAL systems associated with C IV, Si IV, and Mg II ions. The C IV BAL is characterised by a \bi(C~IV) $=15,370$ \kms\ and extends between \vmin(C~IV) $=6,090$ \kms\ and \vmax(C~IV) $=29,500$ \kms\ \citep{Bischetti22}. We identify a Si IV BAL spanning a similar velocity range, with \bi(Si IV) $=7,390$ \kms, and a less prominent Mg II BAL with \bi(Mg~II) $=2,340$ \kms. We cannot assess the presence of a N V BAL in this quasar, because of almost no transmitted flux blueward of Ly-$\alpha$.

We thus classify 12 quasars as HiBALs, and quasar J0923+0402 ($z\simeq6.63$) as a LoBAL. In the case of PSOJ189-15 ($z\simeq5.97$), strong telluric contamination affecting the spectral region blueward of Mg II does not allow us to validate/rule out the presence of a Mg II BAL. Our results increases by a factor of about four the number of Si IV BAL quasars identified at $z\gtrsim6$ \citep{Wang21}. The presence of a Si IV BAL in J2211-3206 ($z=6.33$) was previously suggested by \cite{Chehade18}. A few hints of Mg II absorption in $z\gtrsim6$ quasars have been reported so far \citep[e.g.][]{Maiolino04,Wang21}.

\begin{figure*}[htb]
    \centering
    \includegraphics[width=1\textwidth]{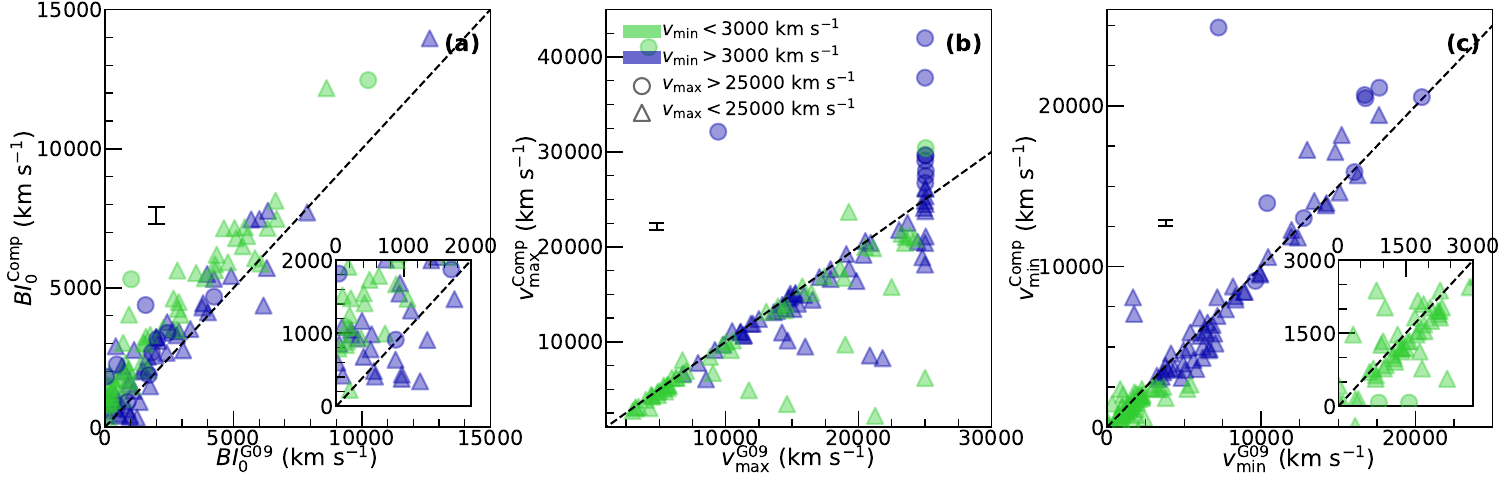}
    \caption{Comparison between the C IV BAL properties measured  with our composite template method,labelled as ``Comp'' (Sect. \ref{subsec:comp}), and in the analysis by \cite{Gibson09} for the quasars in the $2.1<z<5.0$ sample, labelled as ``G09''. (a) balnicity index \bi, (b) maximum velocity \vmax\, and (c) minimum velocity \vmin\ of the BAL outflows. In panel (a) and (c), the inset is a zoom in the region with small \bi\ and \vmin, respectively}. In all panels, the dashed line indicates a 1:1 relation  and the error bar represents the typical uncertainty on the BAL parameters (Sect. \ref{subsec:balprop}).
    \label{fig:gibson-comparison}
\end{figure*}

\begin{figure}[htb]
    \centering
    \includegraphics[width = 1\columnwidth]{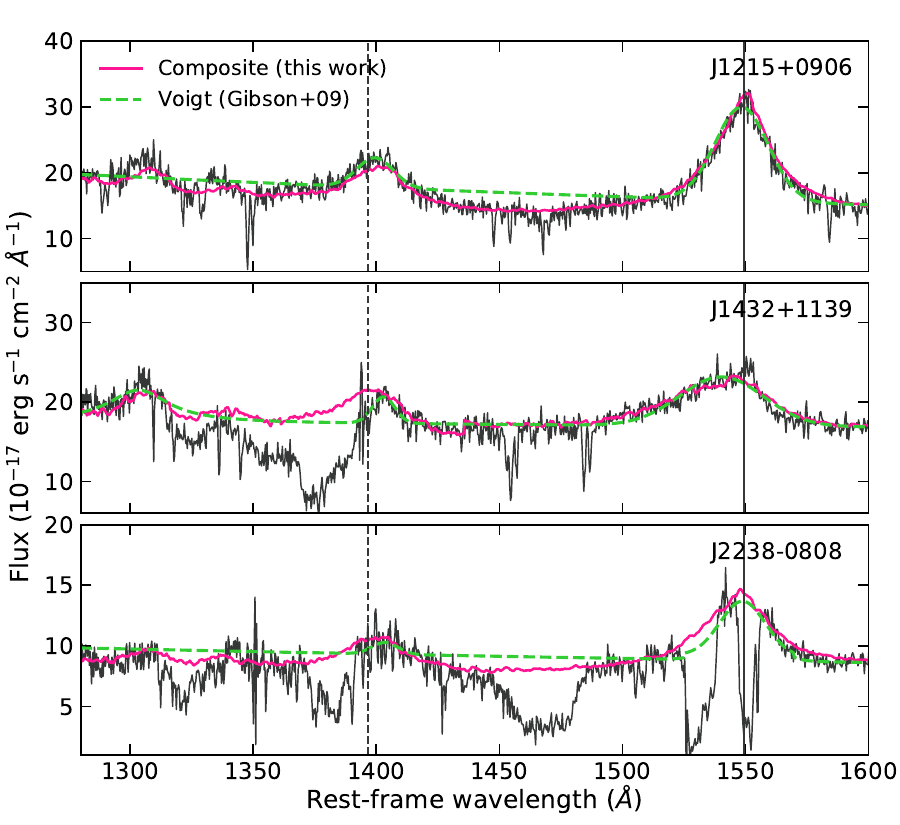}
    \caption{Example of SDSS spectra for three BAL quasars from the $2.1<z<5.0$ sample. We show the comparison between our reconstruction of the intrinsic quasar emission, based on composite template spectra (Sect. \ref{subsec:comp}), and the prescription by \cite{Gibson09}. Different recipes can lead to a different BAL/no-BAL classification (J1215+0906 and J1432+1139), and to different \bi\ (J2238-0808), see Sect. \ref{subsec:sdss} for details. Vertical solid and dashed lines refer to the rest-frame wavelength of C IV and Si IV, respectively.}
    \label{fig:comparison-2}
\end{figure}

The BAL parameters measured for the Si IV and N V BAL outflows are listed in Table \ref{table:otherbal}, and their residual spectra are shown in Fig. \ref{fig:norm_spec}. The Si IV BAL features are characterised by \bi~$\sim200-8,100$ \kms, which are typically smaller (10-60\%) than those measured in C IV, consistently with what has been found for SDSS quasars \citep{Gibson09}. The only exception is PSOJ009-10, for which \bi(C IV) = 1,110 \kms\ \citep{Bischetti22} and \bi(Si IV)=2,880 \kms. However, this quasar shows a very red spectrum with almost no Ly-$\alpha$ and N V emission, overfitted by the composite template at $\lambda<1270$ \AA\ (Fig. \ref{fig:composite-appx}). This suggests strong absorption, consistent with the presence of a Si IV BAL outflow, whose properties are nevertheless highly uncertain. For the N V BALs, we measure \bi~$\sim600-2,300$ \kms.

\subsection{Re-analysis of SDSS quasar spectra}\label{subsec:sdss}

Given that the $2.1<z<5.0$ quasars in our sample did not benefit from a homogeneous identification and characterisation  of BAL outflows from the literature (Sect. \ref{subsec:low-z}), we have re-analysed their spectra following the approach described in Sect. \ref{sect:ident}.
In Fig. \ref{fig:gibson-comparison} we compare our results with those by \cite{Gibson09}, who provided \bi, \vmin\ and \vmax\ for the 1317 SDSS DR5 quasars in our sample \citep{Schneider07}. We identify as BALs 162 out of the 207 BAL quasars found by \cite{Gibson09}. The remaining quasars, classified as non-BAL quasars in our analysis, have only very weak BAL absorption  in \cite{Gibson09} (all but 5 have \bi$<500$ \kms), likely owing to the flat continuum emission in the C IV-Si IV spectral region not well reproduced by the reddened power-law continuum model used by \cite{Gibson09}, introducing a small systematic on the true continuum level. One example is shown in Fig. \ref{fig:comparison-2} top panel for quasar SDSSJ1215$+$0906 ($z=2.72$), previously classified as a BAL quasar with \bi\ $\simeq120$ \kms\ by \cite{Gibson09}, which we instead identify as a non-BAL. In addition, small differences between the DR5 and DR7 spectra can easily produce \bi\ variations of a few hundred \kms\ \citep[e.g.][]{Byun22,Vietri22}. On the other hand, we identify 39 new BAL quasars that were classified as non-BALs by \cite{Gibson09}. One example is SDSS quasar J1432+1139 ($z=2.99$, Fig. \ref{fig:comparison-2} middle panel), showing high-velocity C IV BAL features with \vmin\ $\sim29,000$ \kms, that is outside the spectral range investigated by \cite{Gibson09}.
When comparing the properties of the BAL quasars identified in both our work and \cite{Gibson09}, we find on average no systematic difference in \bi, with a significant scatter at \bi$\lesssim1,000$ \kms\ (Fig. \ref{fig:gibson-comparison}a).   We  measure higher \bi\ for those quasars in which the BAL absorption reaches \vmax $>25,000$ \kms,  as \cite{Gibson09} did not account for absorption troughs beyond this threshold. 
We typically measure larger \bi\, in those cases in which the absorption has \vmin$<3,000$ \kms\ and thus affects the blue side of the C IV line, as it can be seen for SDSS J2238-0808 (z=3.17) in Fig. \ref{fig:comparison-2} bottom panel. This is due to the fact that our composite templates better reproduce the asymmetric C IV profiles, typically showing a more prominent blue wing due to ionised outflows, than the Voigt profiles in \cite{Gibson09}.

The majority of the C IV BAL outflows show  \vmax\ and \vmin\ velocities similar  to those measured by \cite{Gibson09}. In a few cases, we measure lower \vmax\ because of the different continuum treatments, and we identify a subsample of BAL quasars with extremely fast BAL outflows (\vmax$\sim25000-40000$ \kms), missed by \cite{Gibson09}. These high-velocity BAL outflows are visible as a cut-off around $v_{\rm max}^{\rm G09}\sim25,000$ \kms\ in Figure \ref{fig:gibson-comparison}b. Our findings are consistent with the results of \cite{Bruni19, Rodriguez-Hidalgo20}, who reported the presence of BAL outflows reaching velocities of 10-15\% of the light speed in a small fraction of SDSS quasars at $z\sim2-4$. In the case of absorption with \vmin$<3,000$ \kms\ (Fig. \ref{fig:gibson-comparison}c), we typically measure \vmin\ lower by a factor of about two with respect to \cite{Gibson09} estimates, likely because the Voigt profiles employed in their modelling underestimate the intrinsic emission of the C IV blue wing (Fig. \ref{fig:comparison-2} bottom).

\begin{figure}[thb]
    \centering
    \includegraphics[width=1\columnwidth]{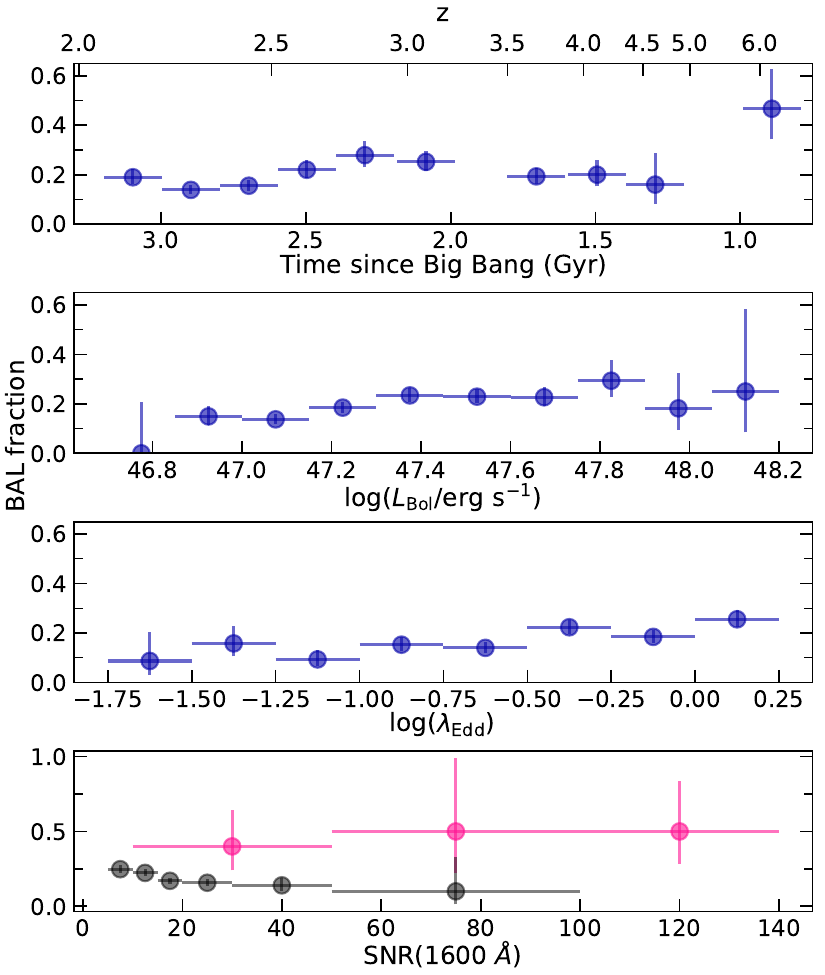}
    \caption{BAL fraction as a function of different physical quantities. Top: cosmic time, from 0.8 to 3.2 Gyr from the Big Bang, in bins of $\sim200$ Myr. Top axis indicate the corresponding redshift interval. Second from top: quasar bolometric luminosity, estimated from the rest-frame optical via bolometric correction. Second from bottom: accretion rate. Bottom: median SNR of the quasar continuum at {rest-frame} 1600 \AA\ (Sect. \ref{sec:balfrac}). Black(magenta) dots refer the $2.1<z<5.0$($5.8<z<6.6$) sample. In each panel, horizontal bars indicate the  bins in which BAL fractions have been calculated, while vertical bars show Poissonian uncertainties at a 90\% confidence level for each bin \citep{Gehrels86}}.
    \label{fig:fbal-z}
\end{figure}

\section{BAL quasar fraction}\label{sec:balfrac}

Figure \ref{fig:fbal-z} shows the evolution of the BAL quasar fraction across a cosmic time interval of about $2.3$ Gyr, corresponding to the redshift range $2.1<z<6.6$ probed by our sample. We find the BAL fraction to be almost constant at $z\lesssim4.5$ with a median value of $\sim20$ \%. The oscillations around this value are likely due to the SDSS quasar selection systematics, affecting differently BAL and non-BAL quasars as a function of redshift, that are minimised but not totally removed by our sample selection (Sect. \ref{subsec:selection}). Indeed, BAL quasars are expected to be less-efficiently selected than non-BALs at $z<2.5$, more efficiently selected at $2.6<z<3.1$, while no significant difference in the selection efficiency is predicted for $3.1<z<3.5$ \citep{Reichard03}. The BAL fraction is instead significantly higher at $z\gtrsim6$, by a factor of about 2.5  \citep{Bischetti22}. The lack of data at $z\sim5$ does not allow us to assess whether the decrease in the BAL fraction at $z<6$ is smooth, consistently with secular evolution of the BH accretion properties, or whether it rapidly drops after $z\sim6$, due to efficient BH feedback at early epochs (see Sect. \ref{subsec:evolution}).

\begin{deluxetable}{lcc}[htb]

\tablenum{3}
\tablecaption{Results from $\chi^2$ and Spearman rank statistical  tests}
\label{table:stats}
\tablewidth{0pt}
\tablehead{
\colhead{Correlation} & $p$-value$_{\chi^2}$ & $p$-value$_S$\\
 & (1) & (2)
}
\startdata
BAL fraction vs. \lbol   & 0.05[0.03-0.08]  &  0.09[0.02-0.52]\\
BAL fraction vs. \ledd   & 0.009[0.002-0.05] & 0.03[0.008-0.11] \\
\vmax\ vs $z$  & $1.3\times10^{-10}$ & $\mathbf{2.0\times10^{-4}}$\\
\vmin\ vs $z$ &  0.025 & 0.016\\
\wmax\ vs $z$ & $6.9\times10^{-5}$ & $\mathbf{9.5\times10^{-4}}$\\
\vmax\ vs \ledd & 0.16[0.06-0.43] &  0.045[0.009-0.24]\\
\vmin\ vs \ledd & 0.81 [0.61-1.0] &  0.45[0.28-0.71]\\
\wmax\ vs \ledd & 0.03 [0.004-0.1] & 0.01[0.002-0.07]\\
\enddata
\tablecomments{ (1) $p-$values associated with the $\chi^2$ difference between a constant and a linear-relation fit. (2)  $p-$values for a Spearman rank's correlation test. Square brackets correspond to the standard deviation on $p$ taking into account the 0.2(0.5)dex uncertainty on \lbol(\ledd)}.
\end{deluxetable} 

To investigate whether the BAL fraction depends on nuclear properties of the quasar, Figure \ref{fig:fbal-z} also shows the BAL fraction as a function of $L_{\rm Bol}$ and $\lambda_{\rm Edd}$.
\cite{Bischetti22} found no significant increase in the BAL fraction of $z\sim2-4$ quasars in the luminosity range log($L_{\rm Bol}$/erg s$^{-1}$)$\simeq 46.5-48$, even after correcting the UV-continuum level for the BAL absorption. However, because BAL quasars typically show redder slopes of the UV continuum \citep[][]{Trump06,Gibson09,Allen11} also in the high-luminosity regime spanned by our sample, using a common bolometric correction may result in an underestimate of \lbol\ in BAL quasars. We do not correct the continuum luminosity for dust extinction because of the limited spectral range covered by the SDSS spectra, resulting in substantial degeneracy between the UV continuum shape and the magnitude of intrinsic reddening \citep[e.g.][]{Gibson09}. Also, one would need to assume an extinction law and possibly also its evolution with redshift \citep[e.g.][]{Gallerani10}, increasing the uncertainties on \lbol. Instead, because all quasars in our sample are detected in the rest-frame optical, we derive an independent estimate of \lbol\ by applying the $\sim5100$ \AA\ bolometric correction by \cite{Runnoe12} to the 2MASS H(K) fluxes for the $2.13<z<3.2$($3.6<z<5.8$) subgroups in our low-z sample, and to the WISE $W1$ fluxes for the high-z sample.  The optical-based and UV-based bolometric luminosities are consistent for non-BAL quasars, while the optical-based \lbol\ are typically higher by a factor of two for BAL quasars. Figure \ref{fig:fbal-z} shows the BAL fraction as a function of the optical-based \lbol. We find a  marginal correlation ($p$-value $\sim0.05\mathbf{-0.09}$, Table \ref{table:stats}) with a BAL fraction increase from $\sim20\%$ to 25\%. A clearer trend ($p$-value $\sim0.009\mathbf{-0.03}$) is observed between the BAL fraction and $\lambda_{\rm Edd}$, with a factor of two increase in the BAL fraction for log$\lambda_{\rm Edd}\in[-1.7,0.2]$ up to $\sim25\%$ although we caution that most of the BH masses for our $2.1<z<5.0$ are based on the C IV line width, which can be significantly altered by the presence of BAL absorption with velocities of a few thousands \kms. This is consistent with the results from previous studies focusing on luminous quasars at $z\sim2-4$ \citep{Dai08,Bruni19}. Nevertheless, neither the \lbol\ nor the $\lambda_{\rm Edd}$ trends can account for the $\sim47\%$ BAL fraction measured for the $5.8<z<6.6$ sample (median log($L_{\rm Bol}/\rm erg\ s^{-1}$)=47.3, median $\lambda_{\rm Edd}= -0.10$). 

\begin{figure}[htb]
    \centering
    \includegraphics[width=0.97\columnwidth]{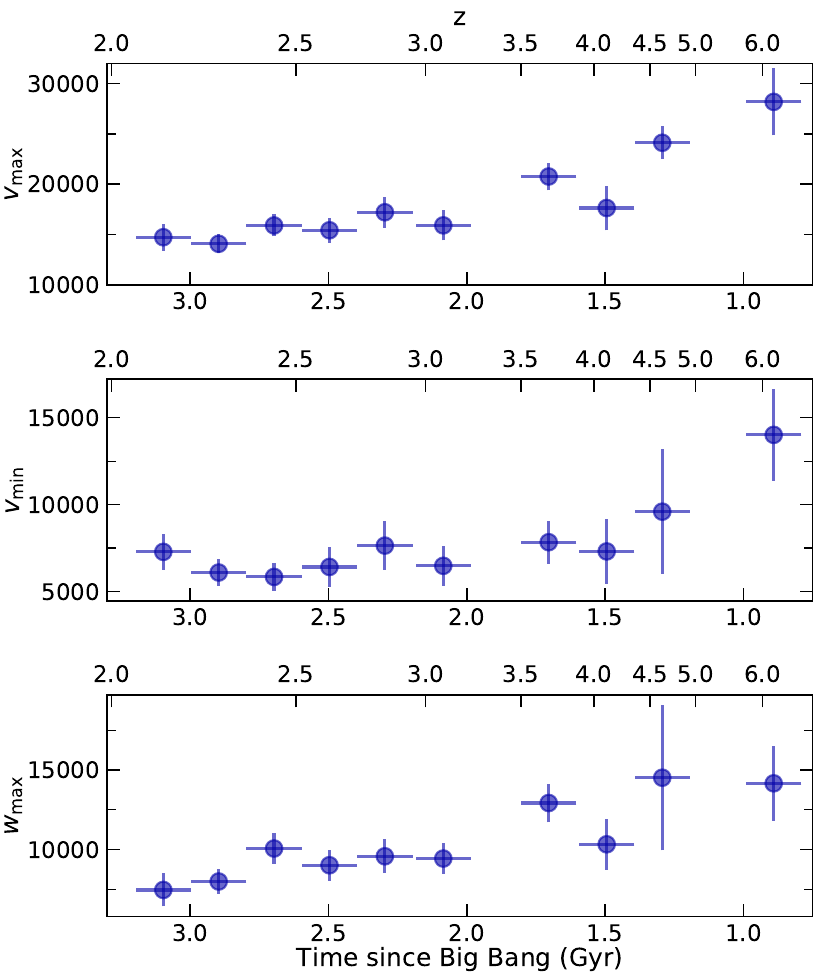}
    \caption{Maximum velocity (top), minimum velocity (middle) and maximum width (bottom) of the C IV BAL outflows, in units of \kms, as a function of cosmic time. Top axes indicate the sampled redshift interval. Horizontal bars indicate the individual time bins of $\sim200$ Myr in which the average BAL velocity has been calculated.  Vertical bars indicate the standard deviation error on the average value for each time bin}. 
    \label{fig:balvel-z}
\end{figure}

\begin{figure*}[htb]
    \centering
\includegraphics[width = 0.9\textwidth]{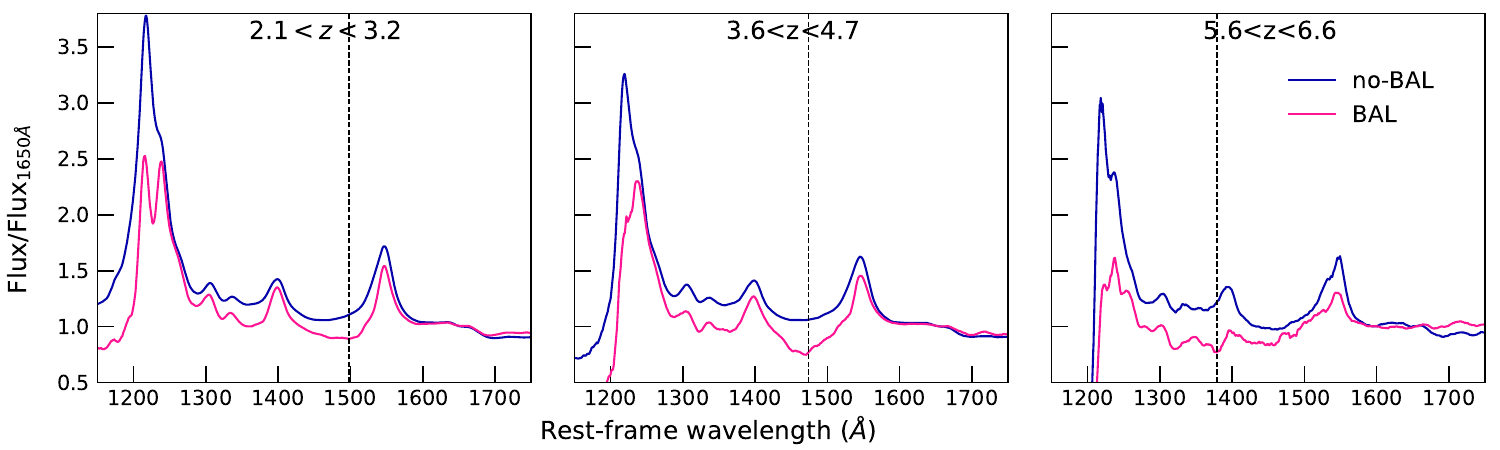}
    \caption{Composite template spectra of the BAL and non BAL quasars in our sample, in three bins of increasing redshift (from left to right). For each panel, composite spectra have been normalised to the median continuum flux at 1600-1700 \AA. The vertical line indicates the rest-frame wavelength corresponding to the maximum depth of the C IV absorption, showing a shift toward lower wavelengths with increasing redshift. In the right panel, we consider only quasars with prominent BAL features \citep[\bi$>1,000$ \kms,][]{Bischetti22}.}
    \label{fig:composite-z}
\end{figure*}

Finally, we investigate the dependence of the BAL fraction on the SNR of the quasar spectra. To this purpose, we homogeneously compute the SNR as the ratio between the median quasar continuum emission in the rest-frame interval 1650-1750 \AA\ and the spectral noise, defined as the median rms in the same spectral region for a 70 \kms\ pixel (Sect. \ref{subsec:comp}).  Figure \ref{fig:fbal-z} (bottom) shows no trend of the BAL with SNR for the $5.8<z<6.6$ sample, while it decreases with increasing SNR for the $2.1<z<5.0$ sample. This is likely due to the combination of (i) a similar rms sensitivity for the spectra, and (ii) bluer continuum slopes, translating into higher SNR for the non-BAL quasars in our sample. Previous studies of BAL quasars in SDSS reported no or little increasing trend (a few \%) of the BAL fraction with SNR, for a similar range of SNR~$=10-50$ \citep{Gibson09, Allen11}, likely reflecting the mild trend with \lbol, due to the wider (by about one order of magnitude) quasar luminosity range sampled by these works. 
We conclude that the increasing trend of the BAL fraction at $z\gtrsim6$ is a genuine cosmic evolution and does not depend either on trends with nuclear properties or on the SNR of the spectra.

\section{Kinematic properties of BAL outflows}\label{sect:kine}
Here we investigate the evolution  of the BAL kinematics across cosmic time, as traced by \vmax, \vmin\ and \wmax\ (Sect. \ref{subsec:balprop}). Top panel of Fig. \ref{fig:balvel-z} shows that the maximum velocity of BAL outflows significantly increases by a factor of about two between $z\sim2-3$ and $z\sim6$  ($p$-value $\mathbf{\lesssim2\times10^{-4}}$, Table \ref{table:stats}), from \vmax$\sim15,000$ to $\sim30,000$ \kms. A similar trend is observed also for the minimum BAL velocity, reaching a typical value of \vmin$\sim15,000$ at $z\sim6$. An increased fraction of BAL quasars with \vmax$\gtrsim30,000$ \kms\ at $z\sim4-4.5$ was recently suggested by \cite{Rodriguez-Hidalgo20}.
Because of the larger increase of \vmax\ with respect to \vmin\ as a function of redshift, the width of the BAL features also increases between $z\sim2-3$ (9,000 \kms) and $z\sim6$ (15,000 \kms).

\begin{figure}[htb]
    \centering
    \includegraphics[width=1\columnwidth]{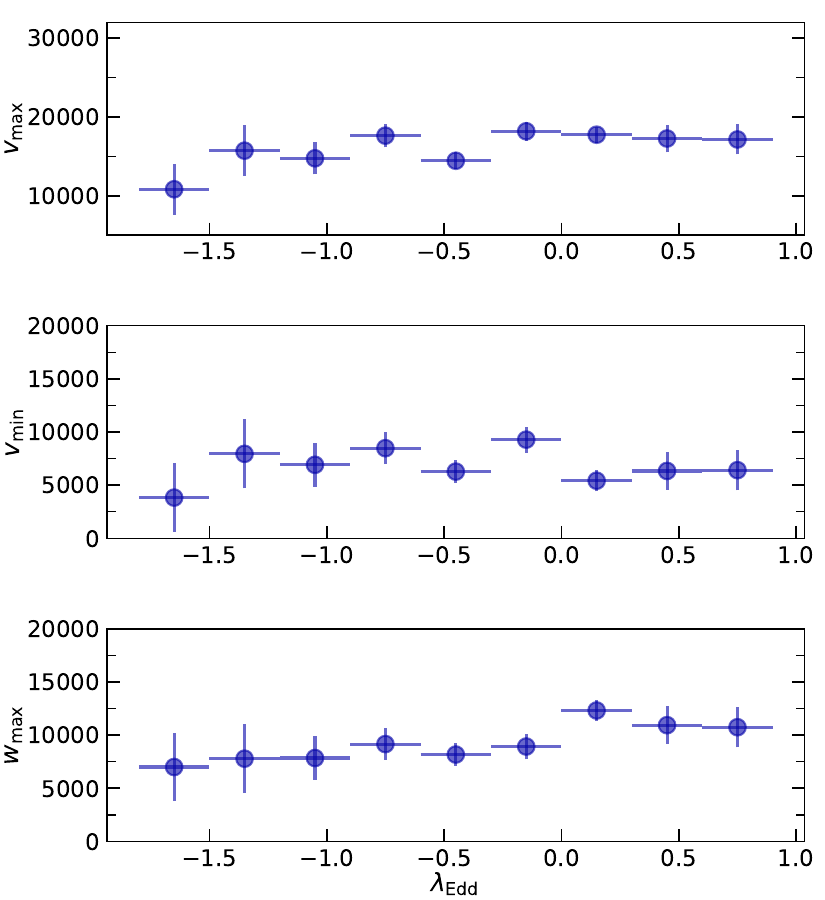}
    \caption{Maximum velocity (top), minimum velocity (middle) and maximum width (bottom) in units of \kms, as a function of $\lambda_{\rm Edd}$ for the C IV BAL outflows in our quasar sample. Horizontal bars indicate the accretion rate bins in which the average BAL velocity has been calculated.  Vertical bars indicate the standard deviation error on the average value for each bin}.
    \label{fig:balvel-edd}
\end{figure}

Figure \ref{fig:composite-z} shows that, because of the combination of the above trends, the composite spectrum of BAL quasars is differently affected by absorption features in three intervals of increasing redshift. The lowest redshift bin ($2.1<z<3.2$) shows that the absorption typically affects the blue side of C IV and the deepest absorption is located at $1,500$ \AA\ (i.e. blueshifted by $\sim9,000$ \kms). At intermediate redshift ($3.6<z<5.0$) the C IV wing is still partially altered by the absorption, which affects the whole spectral range between C IV and Si IV. The deepest absorption is blueshifted by $\sim15,000$ \kms. In the highest redshift interval ($5.8<z<6.6$) the absorption is even broader and extends blueward of Si IV, down to $\sim1,300$ \AA. The deepest absorption is blueshifted by $\sim33,000$ \kms. We note that, owing to the very broad distributions of \vmax, \vmin\, in combination with the small sample size in this high-redshift bin, we include in Fig. \ref{fig:composite-z} only BALs with the most prominent features \citep[\bi$>1,000$ \kms,][]{Bischetti22}. These results, together with the higher BAL fraction observed in quasars at $5.8<z<6.6$, suggest an evolution of the BAL properties with cosmic time. 

To test whether there is a dependence of the BAL kinematics on nuclear quasar properties, we show in Figure \ref{fig:balvel-edd} the relations between \vmax, \vmin\ and \wmax, as a function of the BH accretion rate. We find no evident increase of either \vmax\ and \vmin\ with increasing $\lambda_{\rm Edd}$ over about two orders of magnitude, with an almost constant \vmax$\sim15,000-17,000$ \kms\ and \vmin$\sim5,000-7,000$ \kms\ (Table \ref{table:stats}). There is a hint for a higher \wmax\ ($p$-value $\mathbf{\sim0.01-0.03}$) at high accretion rates, likely due to the lower \vmin\ observed in the three bins with the highest $\lambda_{\rm Edd}$. Similarly, we observe no significant trend of the BAL kinematics with \lbol, likely due to the limited luminosity range spanned by our sample. Indeed, no or only mild trends of increasing \vmax\ with \lbol, $\lambda_{\rm Edd}$ were suggested by previous works studying BAL quasars in SDSS in a wider range of \lbol\ \citep{Ganguly07, Gibson09, Bruni19}. Previous studies at $z\sim2-4$ reported tentative evidence of different properties in BAL quasars with low/high \vmin, in terms of steepness of the X-ray to UV spectral energy distribution or emission line properties, although no significant relation between \vmin\ and quasar luminosity or accretion rate was found \citep{Gibson09,Turnshek88}. We agree with previous results finding no significant correlation between \vmin\ and \ledd\ (Table \ref{table:stats}). These findings confirm that the trend of increasing \vmin, \vmax, and \wmax with redshift in our quasar sample cannot be explained by variations in the quasar nuclear properties, but rather points to an evolutionary effect.

\section{Discussion}\label{sect:discussion}
\subsection{Cosmic evolution of BAL outflow properties}\label{subsec:evolution}

Our results indicate an evolution with redshift of the properties of BAL outflows {in luminous quasars at $2.1<z<6.8$}. We find that the BAL fraction below $z\sim4.5$ is almost constant and close to $20-25\%$ (Fig. \ref{fig:fbal-z}), while it significantly increases up to almost $50\%$ at $z\sim6$. The trend at $2.1<z<4.5$ is consistent with the results of previous studies focusing on rest-frame optical luminous quasar samples \citep{Maddox08, Dai08}, which reported a modest evolution of the BAL fraction with redshift. However, heavily-reddened broad-line quasars are expected to represent a high fraction of the bright quasar population at $z>2$ \citep{Banerji15, Glikman18}, and may be missed by SDSS, from which our $2.1<z<4.5$ sample is drawn.  Several recent surveys \citep[e.g.][]{Schindler19,Boutsia21,Grazian22} have indeed shown that the SDSS selection of bright quasars at high redshift could possibly suffer from incompleteness by up to $30-40\%$ at $z\sim4$, and up to a factor of a few at $z\sim5$ \citep{Wolf20, Onken22}, plausibly affecting also $z\sim6$ quasars. 
Our $5.8<z<6.6$ quasar sample is limited to $J\le19.8-20.0$ magnitudes and probes similar depths in the $J$ and WISE bands \citep[see Table ED1 in ][]{Bischetti22}. Accordingly, we may miss red BAL quasars at $z\sim6$, detected by WISE but too faint in the rest-frame UV to match our selection \citep[e.g.][]{Kato20}. Due to the above limitations and to the limited number of $z\gtrsim5$ quasars in our sample, we cannot accurately probe the BAL fraction evolution in the highest redshift bins. Nevertheless, we can safely assess that the increase in the BAL fraction between $z<4.5$ and $z\sim6$ is a robust evolutionary trend, and that selection effects (e.g. including dust-reddened quasars) would increase this difference even further.

By investigating the dependence of the BAL fraction on quasar nuclear properties, we can safely exclude that the large difference between the BAL fraction at $z\lesssim4.5$ and $z\sim6$ is due to differences in quasar luminosity and accretion rate. Instead, the higher BAL fraction at $z\sim6$ can be explained either with wider angle outflows or as a result of longer blow-out phases of BH-driven outflows compared with $z\lesssim4.5$ \citep{Bischetti22}. A redshift evolution of the BAL geometry is indeed suggested by the increase with redshift of the velocity range spanned by the BAL troughs, as traced by \wmax \citep{Murray98, Elvis00, Hall02}. Although the link of \wmax\ to the velocity dispersion of the outflowing gas is not straightforward, as we mostly observe blends of outflowing components \citep[e.g.][]{Borguet13}, the increase of \wmax\ may be interpreted as stronger turbulence in the higher redshift BAL outflows. In a scenario of Chaotic Cold Accretion (CCA), gas turbulence on scales similar to those of BALs \citep[$\sim$100 pc,][]{Arav18} is linked to the condensation and funnelling of the cold gas phase toward the BH, higher turbulence implying stronger inflows and, in turn, triggering more efficient phases of BH feedback \citep{Gaspari20}.

We find that the BAL velocity (either \vmax\ and \vmin) typically increases with redshift, suggesting that BAL outflows in the high-z Universe might be more easily accelerated to very high velocity than at later cosmic epochs. A viable explanation might be the presence of dust mixed with the ionised gas in the BAL clouds, which would significantly increase the radiation boost efficiency \citep{Ishibashi17, Costa18} and  accelerate the BAL outflows to the observed extreme \vmax$\sim20,000-50,000$ \kms\ \citep{Bischetti22}. At later cosmic epochs, although BAL quasars are known to show relatively redder UV colours \citep[][]{Reichard03, Trump06, Gibson09}, no strong difference in the optical colours of BAL and non-BAL quasars is typically observed, at least in optically-bright sources \citep[e.g.][]{Dai08}. 

Alternative scenarios that could explain the higher BAL velocities at early epochs might be related to different properties of BH accretion at different redshift. CCA toward the BH is expected to be favoured at higher redshift, because of a larger fraction of a clumpy cold gas phase in the quasar hosts, boosting BH-driven outflows to higher velocities \citep{Gaspari17}.
Also, BALs could be launched with higher velocities if BHs are spinning more rapidly at high redshift \citep{King&Pringle08, Zubovas21}.
High BAL velocities have been more commonly found in sources with softer spectral energy distributions, that is lower X-ray to optical luminosity ratios, either from direct X-ray observations of low-z quasars \citep{Laor&Brandt02} or using indirect tracers such as the EW of the He II $\lambda1640$ \AA\ emission line in SDSS quasars, \citep{Richards11,Baskin15,Rodriguez-Hidalgo20}. For the highest redshift bins in which we observe the largest values of \vmax, \vmin\ (see Fig. \ref{fig:balvel-z}), there is very limited information on the X-ray properties. Only a few tens of quasars have been targeted with sensitive observations at $z>5$, including only a few BAL quasars, some of which do exhibit the softest optical-to-X-ray slopes  \citep{Nanni17, Vito19, Wang20}. The weakness of the He II emission line makes its detection very challenging at these redshifts \citep[e.g.][]{Shen19}. The characterisation of the He II emission in the $5.8<z<6.6$ sample, enabled by the high-SNR of the \xshoot\ spectra from the \xqr\ survey, will be presented in a forthcoming paper.

\subsection{Impact of BAL outflows on BH and galaxy-evolution} 

The broad and often saturated BAL profiles associated with CIV prevent us from an accurate measurement of the outflowing gas mass and hence of the energy injected by the outflow into the galaxy medium \citep{Dunn12,Borguet13}. Nevertheless, assuming that BAL outflow masses at $z\sim6$ are similar to those measured using unsaturated absorption lines in $z\sim2$ quasars  \citep{Moe09,Dunn10}, we can estimate that BAL quasars at $z\sim6$ globally inject about 20 times more energy with respect to $z\sim2-4$ quasars \citep{Bischetti22}. Indeed,  the outflow kinetic power linearly scales with the BAL fraction and with the third power of \vmax\ \citep[Eq. 5 and 6 in ][]{Bischetti17}. This result strongly points towards a phase of efficient BH feedback occurring at $z\sim6$, as the energy injected by these BAL outflows will likely suppress gas accretion and slow down BH growth \citep[e.g][]{Torrey20}. This likely represents the first observational evidence of what has been so far only predicted by cosmological simulations of the early assembly of bright quasars. BHs at the centre of bright quasars are indeed expected to grow exponentially at very high redshift ($z>>6$), with accretion rates close to the Eddington limit or beyond \citep{Inayoshi16,Pezzulli16}, while around $z=6$ BH growth is expected to significantly slow down because BH feedback has become strong enough to remove gas from the central galaxy regions and to prevent further accretion onto the BH \citep{Costa14,vanderVlugt19}. BHs powering bright quasars at $z\sim6$, with typical masses of $\sim10^9$ M$_\odot$ \citep{Mazzucchelli17,Shen19,Farina22}, have been found to be typically over-massive by a factor of $\sim10$ with respect to the mass of their host-galaxies, when compared to the BH mass vs galaxy dynamical mass relation observed in the low-redshift Universe \citep{Gaspari19,Pensabene20,Neeleman21}. State-of-the-art measurements of the BH mass (based on the Mg II line), and of the dynamical mass (based on spatially-resolved C II $\lambda158$ $\mu$m observations) for the quasars in our $5.8<z<6.6$ sample confirm this result, indicating that BH growth must have dominated over the host-galaxy growth at $z>>6$ and that a transition epoch during which BH growth decelerates must occur at $z\lesssim6$, leading toward the symbiotic BH and galaxy growth \citep[e.g.][]{Lamastra10,Zubovas21,Inayoshi22}. An early BH mass assembly in bright quasars occurring at $z>6$ is also supported by the fact that the largest BH masses measured at $z\sim6$ and at $z\sim2$ only differ by a factor of a few \citep{Trakhtenbrot21} and by the increase of the median BH accretion rate with redshift for a given quasar luminosity \citep{Yang21,Farina22}. 

The above scenario points toward BH feedback as a likely driver of BH growth suppression, but other physical processes might be in place. The cosmic evolution of the BAL fraction is key to identify the dominant mechanism responsible for slowing down BH growth. A sharp decrease of the BAL fraction occurring between $z\sim6$ and $z\sim5$ would strongly point toward a phase of efficient BH feedback occurting on a short timescale \citep[few tens of Myrs][]{Negri17} and quickly removing gas from the central regions of the galaxy \citep{Zubovas&King13}. Conversely, a smooth decrease of the BAL fraction from $z\sim6$ to $z\sim4$, would rather indicate that several processes may be at play in the transition, including a (less efficient) BH feedback and secular processes, such as a change in the merger rate or in the gas condensation rate, reducing the active phase duty cycle on a timescale of several hundreds of Myrs \citep{Gaspari19,O'Leary21}. To discriminate between these competing scenarios, a crucial step is to obtain a reliable measure of the evolution of the BAL fraction with redshift in representative samples, with a cosmic time sampling $<100-200$  Myr. Currently, the sample presented in this work and, particularly, the lack of sources at $5<z<5.8$ do not allow us to accurately probe the high-redshift evolution of the BAL fraction between $z\sim4.5$ and $z\sim6$. New high quality spectroscopic observations of an absorption-unbiased sample of rest-frame optical bright quasars are necessary to fill this gap and test whether efficient BH feedback is the dominant mechanism leading to the symbiotic BH-galaxy growth phase.

On the other hand, it is unclear whether and on what timescales BAL outflows can affect the evolution of the host-galaxies of $z\sim6$ quasars. Observations of BAL quasars at intermediate redshift report mass outflow rates up to several hundreds of M$_\odot$/yr \citep[e.g][]{Fiore17,Bruni19}, consistent with theoretical expectations from models of BH-driven outflows for an efficient feedback mechanism \citep{Faucher12,Zubovas&King12} and outflow sizes based on photo-ionisation analysis that can reach kpc scales \citep{Arav18,Byun22}, suggesting that BAL outflows may also affect the physical properties of the galaxy medium and, in turn, the galaxy growth. 

 If the effect of BAL feedback extends to kpc scales, host-galaxy properties such as the star formation rate (SFR) would be expected to be different in BAL and non-BAL quasars. If conflicting results have been reported at $z\sim2$ \citep[e.g.][]{Zhang14,Wethers20}, and at $z\sim6$ large uncertainties affect SFR measurements \citep{WangF19, Tripodi22,DiMascia23} and do not allow us to investigate  differences within our $5.8<z<6.6$ sample.
A correlation is also expected between the strength of BH feebdack and the cooling rate of the hot and warm gas phases in the quasar host galaxies, due to heating and turbulence injection into the ISM \citep[e.g.][]{Gaspari15}. 

\section{Conclusion}\label{sec:conclusion}
 
In this work we analyse a sample of 1935 luminous (bolometric luminosity $L_{\rm bol}\gtrsim10^{46.5}$ erg s$^{-1}$) quasars at $z=2.1-6.6$, drawn from the Sloan Digital Sky Survey \citep{Shen11} and from the X-shooter legacy survey of Quasars at Reionisation \citep{Bischetti22}, to investigate the evolution with cosmic time of the BAL fraction and of the kinematics of BH-driven outflows, as traced by BAL features.
Targeting rest-frame optical bright quasars allows us to to reduce biases due to quasar selection criteria (Sect. \ref{subsec:selection}). 

We apply a homogeneous BAL identification method to the total sample, based on composite template spectra to estimate the intrinsic quasar continuum and line emission (Sect. \ref{sect:ident}). This approach allows us to well reproduce the spectral region between Ly-$\alpha$ and C IV for a variety of BAL shapes, without any assumptions on the continuum shapes, nor on the spectral regions in which absorption might occur. At the same time, it takes into account the asymmetry often observed in the C IV profile due to the presence of outflowing gas. 

We find that the BAL fraction is about 20\% and does not vary strongly across this redshift range, in agreement with previous works \citep{Dai08,Maddox08}, while it increases to almost 50\% at $z\sim6$ \citep{Bischetti22}. We also investigate the dependence of the BAL fraction with quasar nuclear properties such as \lbol, $\lambda_{\rm Edd}$, and we observe only weak correlations given the ranges of luminosity and accretion rate probed by our sample, in agreement with previous results for SDSS quasars \citep[e.g.][]{Gibson09,Bruni19}. These trends cannot account for the increase in the BAL fraction observed in the $5.8<z<6.5$ sample (Sect. \ref{sec:balfrac}).

We also observe a redshift evolution of the BAL kinematics. Both \vmax\ and \vmin\ increases at $z\gtrsim4$, the typical BAL velocities at $z\sim6$ being a factor of 2-3 higher than what observed at $z<4$ (Sect. \ref{sect:kine}). The width of the BAL features also likely increases at $z\gtrsim4$. These trends suggest a possible evolution of the BAL geometry and are consistent with BALs being more easily accelerated at early cosmic epochs (Sect. \ref{subsec:evolution}). By investigating the dependence of the BAL kinematics with \lbol, $\lambda_{\rm Edd}$, we were able to exclude that the redshift evolution is due to different luminosity and/or accretion properties within the sample.
 
BAL outflows being more common and faster at $z\sim6$ imply that strong BH feedback is likely occurring around this epoch, owing to the injection of large amounts of energy into the BH surroundings and in the galaxy medium, in agreement with expectations from galaxy evolution models \citep{vanderVlugt19,Inayoshi22}. However, the limited number of high-redshift quasars in our sample does not allow us to accurately probe the $z\gtrsim4.5$ evolution of the BAL fraction on a timescale of 100-200 Myr. This hampers us from discriminating between a sudden or a smooth change of the BAL fraction with increasing redshift and, in turn, from assessing whether BH feedback is  driving this evolution. Building a larger, absorption-unbiased sample of rest-frame optical bright quasars with high-quality optical and near-IR spectroscopy will be fundamental to observationally quantify the impact of BH feedback on early BH growth. By complementing this sample with high-frequency ALMA and JWST observations we will be able to measure the host-galaxy growth and to assess whether quasar feedback at $z\sim6$ drives the onset of the symbiotic BH-and-galaxy evolution observed in the lower redshift Universe.

\begin{acknowledgments}
Acknowledgements: this work is based on observations collected at the European Organisation for Astronomical Research in the Southern Hemisphere under ESO large programme 1103.A-0817(A). Funding for SDSS and SDSS-II has been provided by the Alfred P. Sloan Foundation, the Participating Institutions, the National Science Foundation, the US Department of Energy, the National Aeronautics and Space Administration, the Japanese Monbukagakusho, the Max Planck Society and the Higher Education Funding Council for England. The SDSS website is http://www.sdss.org/. The SDSS is managed by the Astrophysical Research Consortium for the Participating Institutions. The Participating Institutions are the American Museum of Natural History, the Astrophysical Institute Potsdam, the University of Basel, the University of Cambridge, Case Western Reserve University, the University of Chicago, Drexel University, Fermilab, the Institute for Advanced Study, the Japan Participation Group, Johns Hopkins University, the Joint Institute for Nuclear Astrophysics, the Kavli Institute for Particle Astrophysics and Cosmology, the Korean Scientist Group, the Chinese Academy of Sciences (LAMOST), Los Alamos National Laboratory, the Max-Planck-Institute for Astronomy, the Max-Planck-Institute for Astrophysics, New Mexico State University, Ohio State University, the University of Pittsburgh, the University of Portsmouth, Princeton University, the United States Naval Observatory, and the University of Washington. M.B., C.F., and F.F. acknowledge support from the PRIN MIUR project ‘Black hole winds and the baryon life cycle of galaxies: the stone-guest at the galaxy evolution supper’, contract number 2017PH3WAT, and support from INAF under PRIN SKA/CTA 2016 FORECaST project. MG acknowledges partial support by HST GO-15890.020/023-A and the {\it BlackHoleWeather} program. K.Z. acknowledges support by Research Council Lithuania grant no. S-MIP-20-43. E.P.F. is supported by the international Gemini Observatory, a program of NSF’s NOIRLab, which is managed by the Association of Universities for Research in Astronomy (AURA) under a cooperative agreement with the National Science Foundation, on behalf of the Gemini partnership of Argentina, Brazil, Canada, Chile, the Republic of Korea, and the United States of America.
\end{acknowledgments}

%

\facilities{VLT(UT2 X-shooter), Sloan}


\software{python 3.8 \citep{Python3}, numpy \citep{Numpy20}, astropy \citep{Astropy13,Astropy18}, pandas \citep{Pandas10,Pandas20}, matplotlib \citep{Matplotlib}
          }




\bibliography{balbib}{}
\bibliographystyle{aasjournal}

\appendix

\begin{figure*}[thb]
    \centering
    \includegraphics[width = 0.9\textwidth]{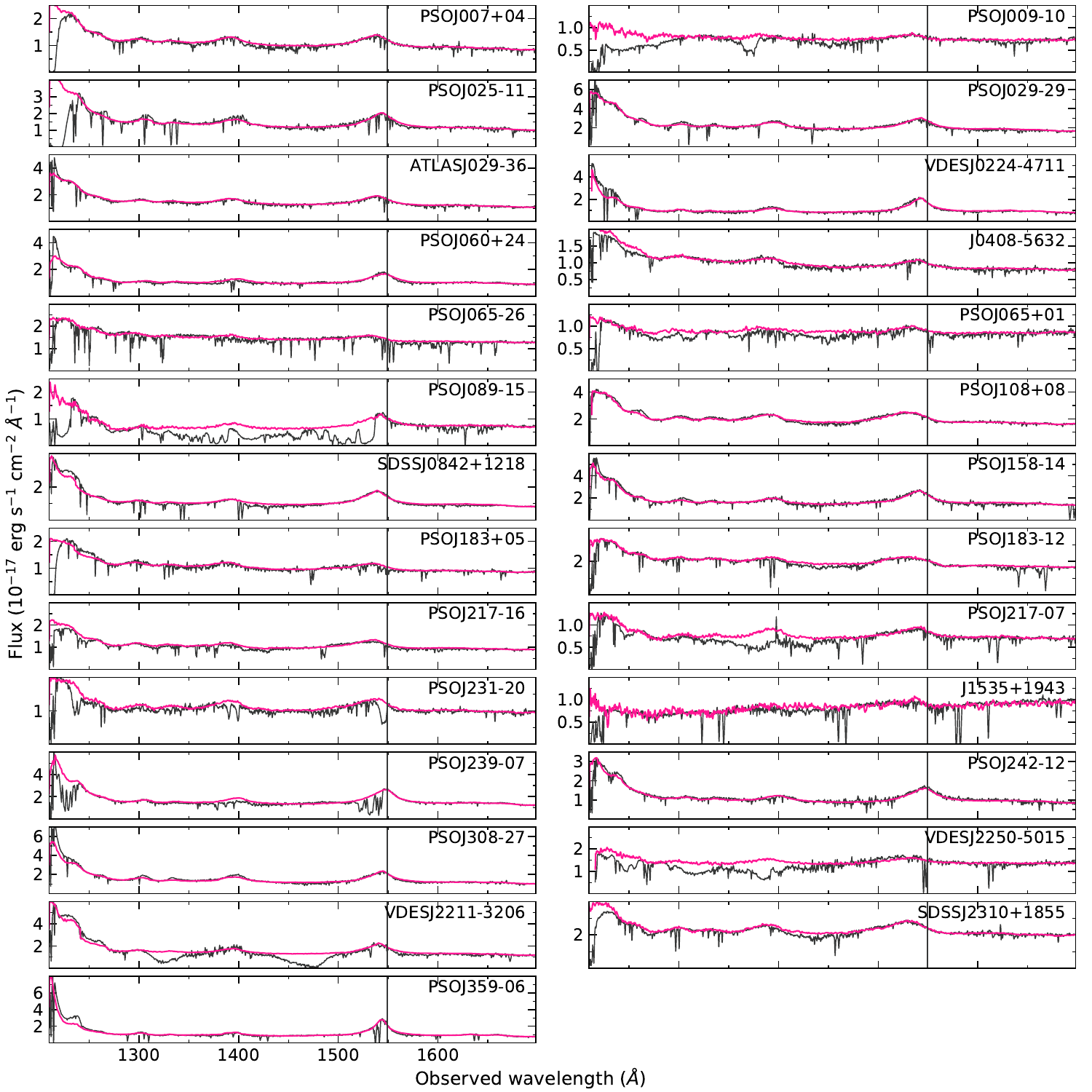}
    \caption{X-Shooter spectra of the $5.8<z<6.6$ quasars in our sample (black curves), binned to 150 \kms. Composite templates, used to estimate the intrinsic quasar
emission are shown by the magenta curve. The vertical line indicates the rest-frame wavelength of C IV.}
    \label{fig:composite-appx}
\end{figure*}

\begin{figure*}[htb]
    \centering
    \includegraphics[width=0.4\textwidth]{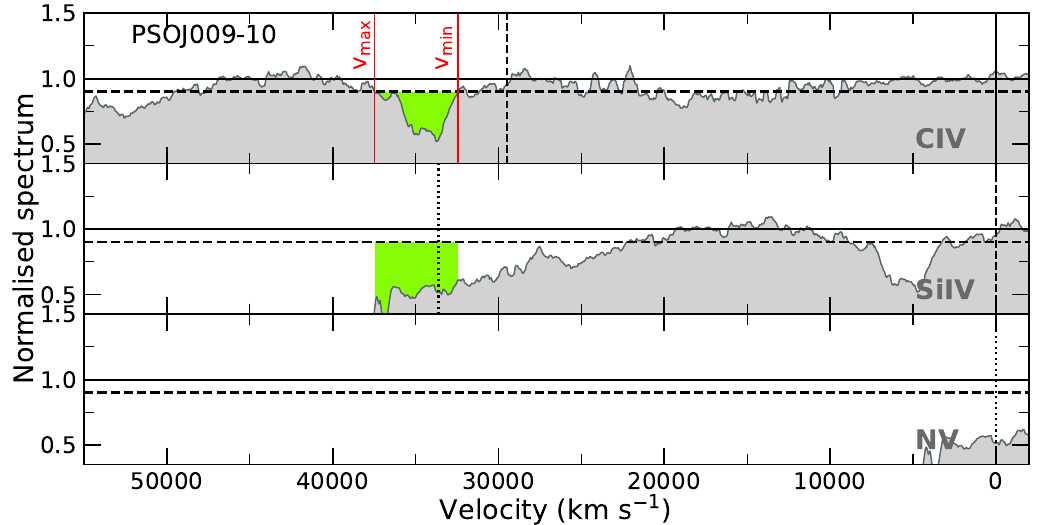}
    \includegraphics[width=0.4\textwidth]{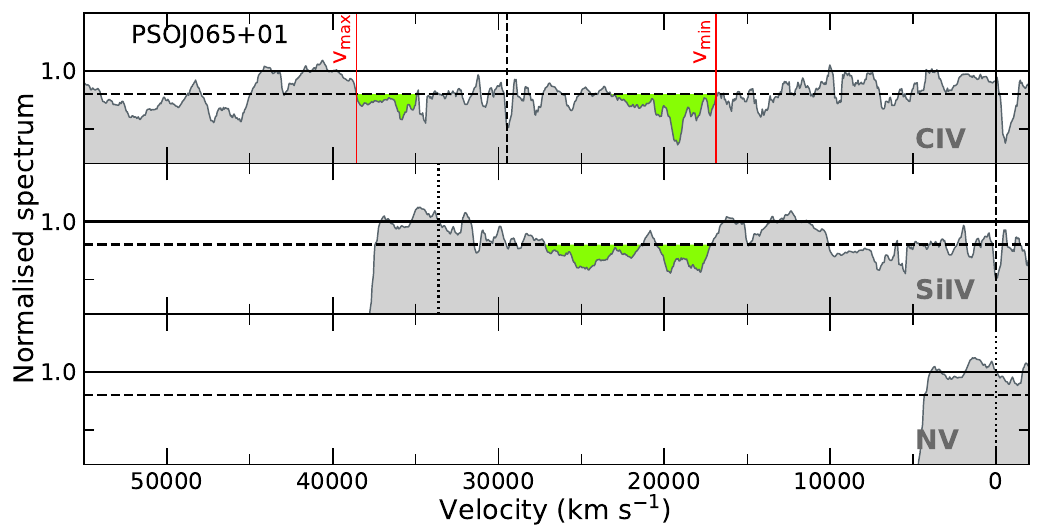}\vspace{0.15cm}
    \includegraphics[width=0.4\textwidth]{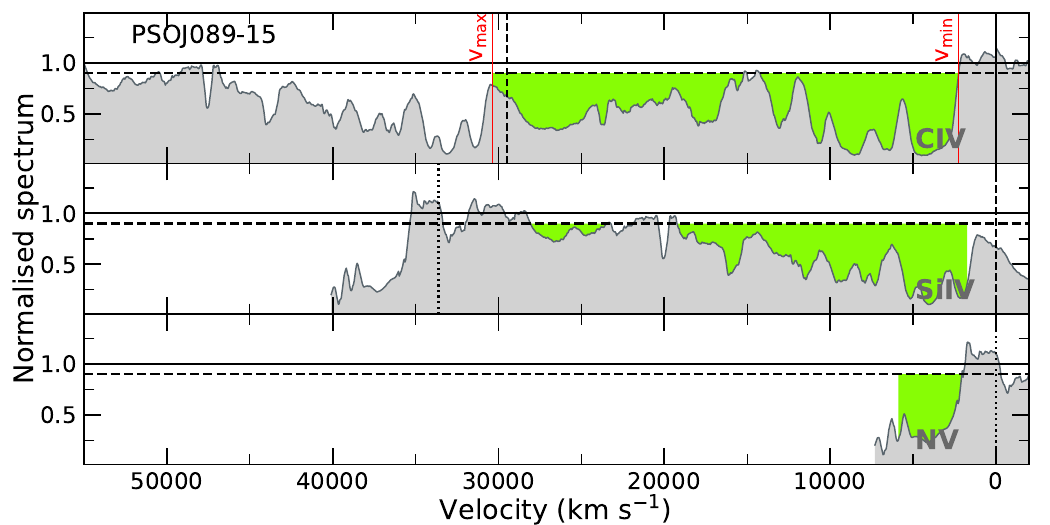}
    \includegraphics[width=0.4\textwidth]{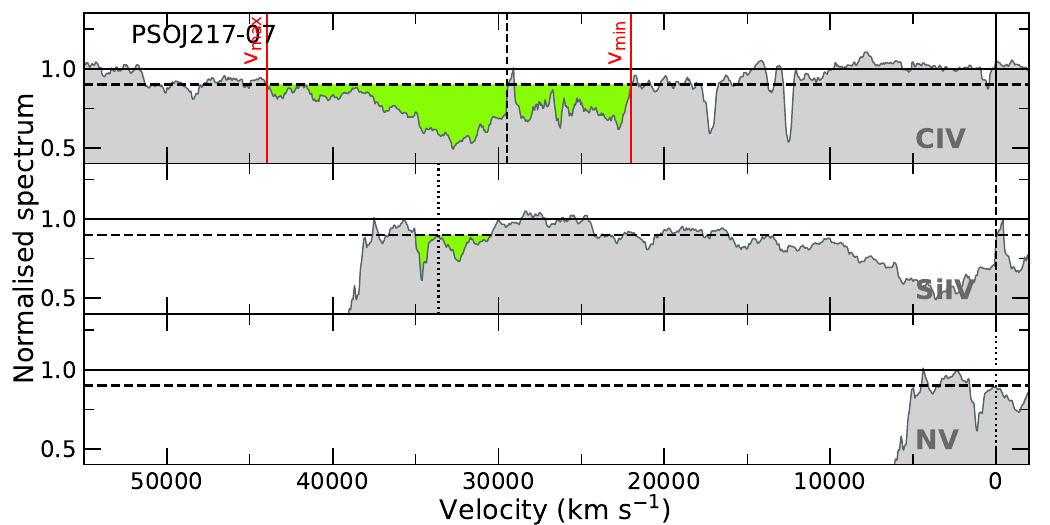}\vspace{0.15cm}
    \includegraphics[width=0.4\textwidth]{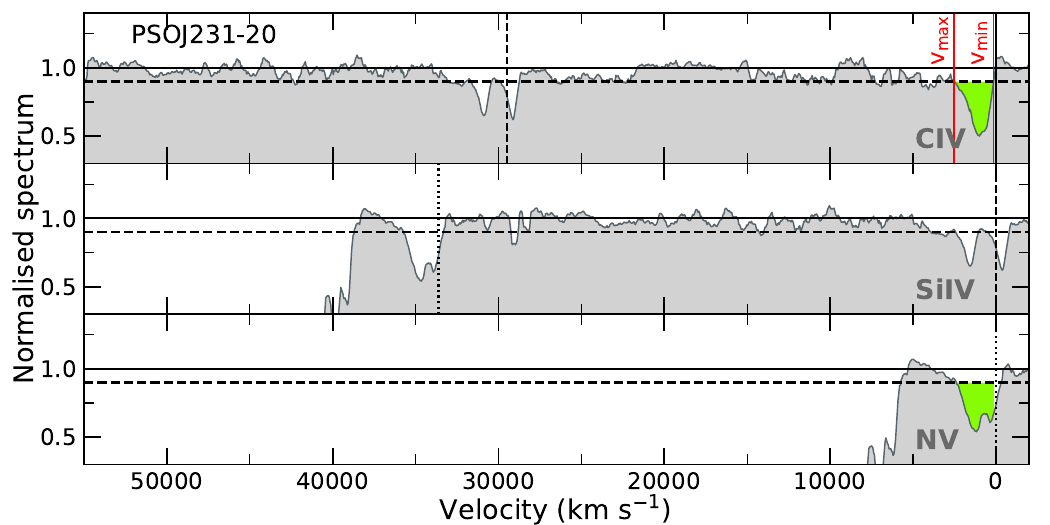}
    \includegraphics[width=0.4\textwidth]{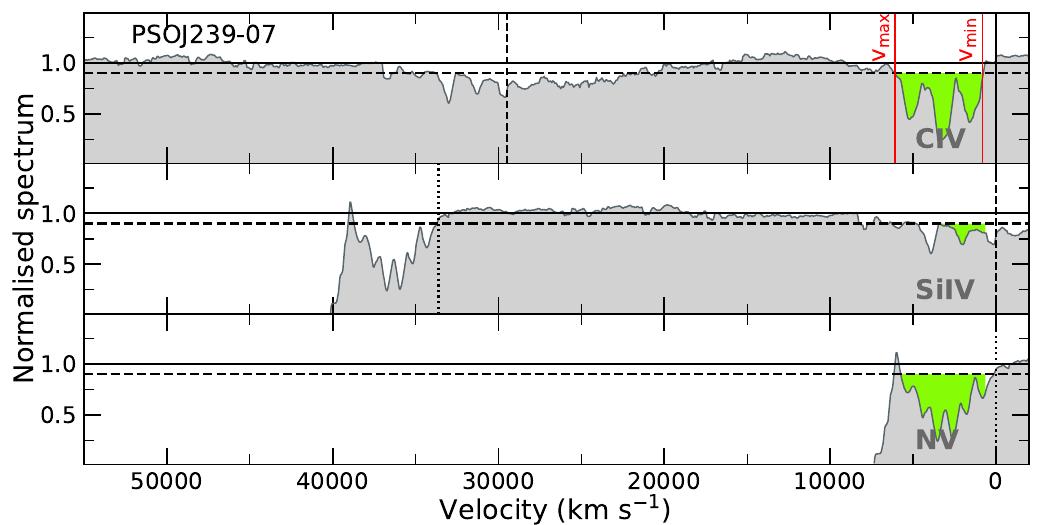}\vspace{0.15cm}
    \includegraphics[width=0.4\textwidth]{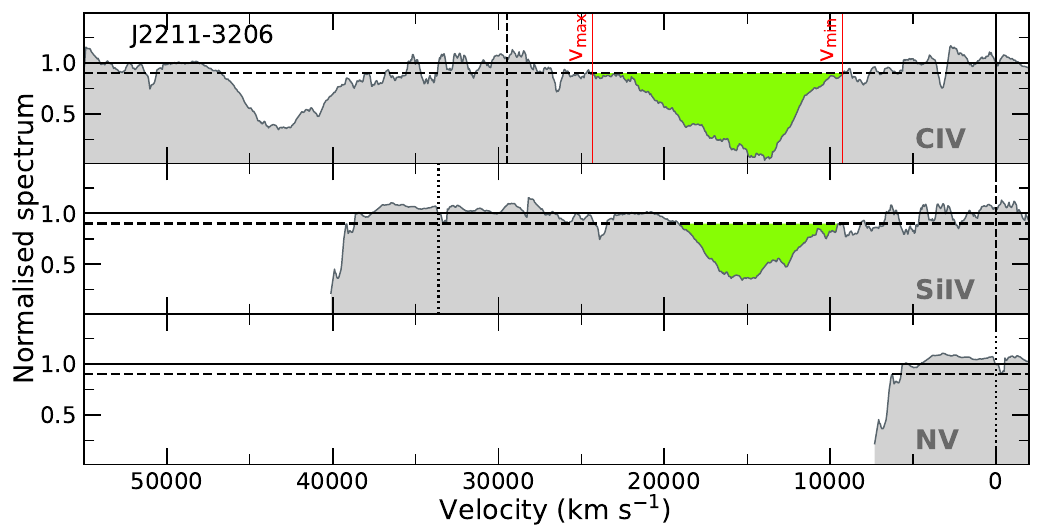}
    \includegraphics[width=0.4\textwidth]{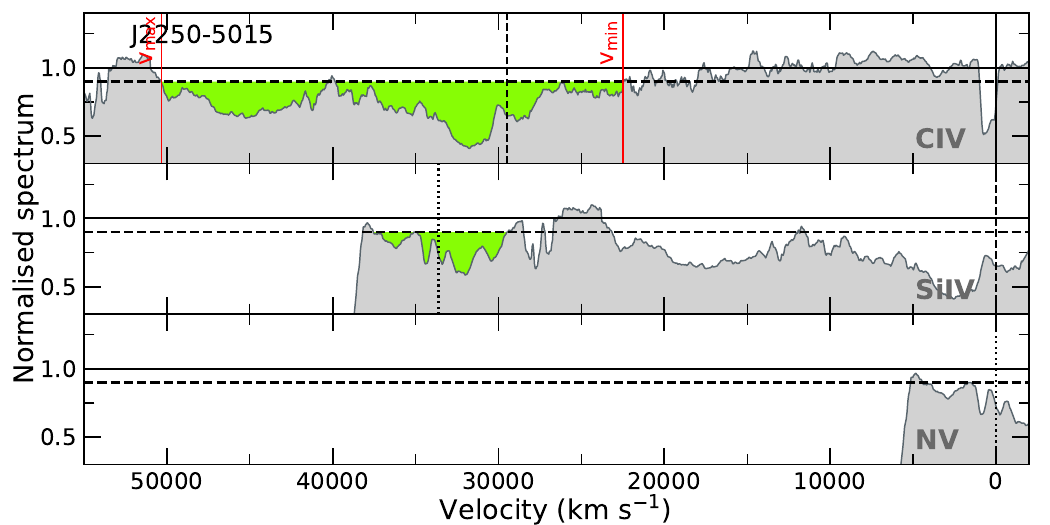}
    \caption{Normalised spectra for the BAL quasars identified in the $5.8<z<6.6$ sample, smoothed to 500 \kms.  The velocity axis in each panel is relative to the rest-frame wavelength of the ionic species indicated by the label. Vertical solid, dashed, and dotted lines indicate the velocity associated with C IV, Si IV, and N V emission lines, respectively. The solid(dashed) horizontal line represents a flux level of 1.0(0.9). BAL troughs, corresponding to a flux level $<0.9$ (Eq. \ref{eq:bi}), are highlighted as green shaded areas.}
    \label{fig:norm_spec}
\end{figure*}



\end{document}